\documentstyle[12pt,psfig]{article}

\catcode`\@=11
\long\def\@makefntext#1{
\protect\noindent \hbox to 3.2pt {\hskip-.9pt
$^{{\ninerm\@thefnmark}}$\hfil}#1\hfill}		

\def\@makefnmark{\hbox to 0pt{$^{\@thefnmark}$\hss}}  

\def\ps@myheadings{\let\@mkboth\@gobbletwo
\def\@oddhead{\hbox{}
\rightmark\hfil\ninerm\thepage}
\def\@oddfoot{}\def\@evenhead{\ninerm\thepage\hfil
\leftmark\hbox{}}\def\@evenfoot{}
\def\sectionmark##1{}\def\subsectionmark##1{}}

\renewcommand{\thefootnote}{\fnsymbol{footnote}}

\newcounter{sectionc}\newcounter{subsectionc}\newcounter{subsubsectionc}
\renewcommand{\section}[1] {\vspace*{0.6cm}\addtocounter{sectionc}{1}
\setcounter{subsectionc}{0}\setcounter{subsubsectionc}{0}\noindent
	{\normalsize\bf\thesectionc. #1}\par\vspace*{0.4cm}}
\renewcommand{\subsection}[1] {\vspace*{0.6cm}\addtocounter{subsectionc}{1}
	\setcounter{subsubsectionc}{0}\noindent
	{\normalsize\it\thesectionc.\thesubsectionc. #1}\par\vspace*{0.4cm}}
\renewcommand{\subsubsection}[1]
{\vspace*{0.6cm}\addtocounter{subsubsectionc}{1}
	\noindent {\normalsize\rm\thesectionc.\thesubsectionc.\thesubsubsectionc.
	#1}\par\vspace*{0.4cm}}

\newcounter{appendixc}
\newcounter{subappendixc}[appendixc]
\newcounter{subsubappendixc}[subappendixc]

\renewcommand{\appendix}[1] {\vspace*{0.6cm}
        \refstepcounter{appendixc}
        \setcounter{figure}{0}
        \setcounter{table}{0}
        \setcounter{equation}{0}
        \renewcommand{\thefigure}{\Alph{appendixc}.\arabic{figure}}
        \renewcommand{\thetable}{\Alph{appendixc}.\arabic{table}}
        \renewcommand{\theappendixc}{\Alph{appendixc}}
        \renewcommand{\theequation}{\Alph{appendixc}.\arabic{equation}}
        \noindent{\bf Appendix \theappendixc #1}\par\vspace*{0.4cm}}

\def\abstracts#1{{

\centering{\begin{minipage}{12.2truecm}\footnotesize\baselineskip=12pt\noindent
	\centerline{\footnotesize ABSTRACT}\vspace*{0.3cm}
	\parindent=0pt #1
	\end{minipage}}\par}}

\renewenvironment{thebibliography}[1]
	{\begin{list}{\arabic{enumi}.}
	{\usecounter{enumi}\setlength{\parsep}{0pt}
\setlength{\leftmargin 1.25cm}{\rightmargin 0pt}
	 \setlength{\itemsep}{0pt} \settowidth
	{\labelwidth}{#1.}\sloppy}}{\end{list}}

\newcounter{itemlistc}
\newcounter{romanlistc}
\newcounter{alphlistc}
\newcounter{arabiclistc}

\newcommand{\fcaption}[1]{
        \refstepcounter{figure}
        \setbox\@tempboxa = \hbox{\footnotesize Fig.~\thefigure. #1}
        \ifdim \wd\@tempboxa > 6in
           {\begin{center}
        \parbox{6in}{\footnotesize\baselineskip=12pt Fig.~\thefigure. #1}
            \end{center}}
        \else
             {\begin{center}
             {\footnotesize Fig.~\thefigure. #1}
              \end{center}}
        \fi}

\newcommand{\tcaption}[1]{
        \refstepcounter{table}
        \setbox\@tempboxa = \hbox{\footnotesize Table~\thetable. #1}
        \ifdim \wd\@tempboxa > 6in
           {\begin{center}
        \parbox{6in}{\footnotesize\baselineskip=12pt Table~\thetable. #1}
            \end{center}}
        \else
             {\begin{center}
             {\footnotesize Table~\thetable. #1}
              \end{center}}
        \fi}

\def\@citex[#1]#2{\if@filesw\immediate\write\@auxout
	{\string\citation{#2}}\fi
\def\@citea{}\@cite{\@for\@citeb:=#2\do
	{\@citea\def\@citea{,}\@ifundefined
	{b@\@citeb}{{\bf ?}\@warning
	{Citation `\@citeb' on page \thepage \space undefined}}
	{\csname b@\@citeb\endcsname}}}{#1}}

\newif\if@cghi
\def\cite{\@cghitrue\@ifnextchar [{\@tempswatrue
	\@citex}{\@tempswafalse\@citex[]}}
\def\citelow{\@cghifalse\@ifnextchar [{\@tempswatrue
	\@citex}{\@tempswafalse\@citex[]}}
\def\@cite#1#2{{$\null^{#1}$\if@tempswa\typeout
	{IJCGA warning: optional citation argument
	ignored: `#2'} \fi}}

 1
 1
 1

\font\ninerm=cmr9

\def\la{\langle}
\def\ra{\rangle}
\def\a{\alpha}
\def\b{\beta}
\def\d{\delta}
\def\o{\omega}

\begin{document}
\begin{flushright}
Utrecht THU-95/23\\
hep-th/9510026
\end{flushright}

\centerline{\normalsize\bf INTRODUCTION TO BLACK HOLE MICROSCOPY}
\vfill
\vspace*{0.6cm}
\centerline{\footnotesize THEODORE A. JACOBSON}
\baselineskip=13pt
\centerline{\footnotesize\it Department of Physics, University of Maryland}
\baselineskip=12pt
\centerline{\footnotesize\it College Park, MD 20742-4111 USA}
\centerline{\footnotesize E-mail: jacobson@umdhep.umd.edu}
\centerline{\footnotesize and}
\centerline{\footnotesize\it Institute for Theoretical Physics, University of
Utrecht}
\baselineskip=12pt
\centerline{\footnotesize\it P.O. Box 80.006, 3508 TA Utrecht, The Netherlands}

\vspace*{0.9cm}
\abstracts{The aim of these notes is both to review the standard
understanding of the Hawking effect, and to discuss the modifications to this
understanding that might be required by new physics at short distances.
The fundamentals of the Unruh effect are reviewed, and
then the Hawking effect is explained as a ``gravitational Unruh effect",
with particular attention to the state-dependence of this picture.
The order of magnitude of deviations from the thermal spectrum of
Hawking radiation is estimated under various hypotheses on physics at
short distances. The behavior of black hole radiation
in a linear model with altered short distance
physics---the Unruh model---is discussed in detail.}

\normalsize\baselineskip=15pt
\setcounter{footnote}{0}
\renewcommand{\thefootnote}{\alph{footnote}}
\section{Introduction}

When a black hole forms in a collapse process, virtually all features
of the original collapsing object are erased. All that survives of its
origin is the total mass, angular momentum, and charges. What remains
is a pure object, made of nothing but empty spacetime. Thus a non-rotating
uncharged black hole is characterized by its Schwarzschild radius
$R_S=2GM/c^2$. The length $R_S$ provides the only scale.

If we bring in $\hbar$ something new happens. The ``empty space" that the
black hole is made of wakes up and is full of vacuum fluctuations of all fields
(including, presumably, the gravitational field itself). And these vacuum
fluctuations are propagating in the black hole background.
Hawking discovered\cite{Hawking} in 1974 that this black hole vacuum is not
stable.
He showed that a thermal flux of positive energy at temperature
$T_H=\hbar c/4\pi R_S$ flows away from the vicinity of the black hole.
(For a stellar mass black hole this temperature is very low, since the
Schwarzschild radius is about a kilometer
($T_H\simeq 10^{-7}{\rm K}\, (1.5{\rm km}/R_S)$), but the temperature
could be very high for a much smaller black hole that might have formed in the
early universe.)
Moreover, negative energy flows across the horizon into the hole, so total
energy in the fixed, static, background is conserved. Presumably the
gravitational back-reaction to this process diminishes the mass of the
hole, and the process continues until the hole has lost all of its mass.

The fact that the Hawking radiation has a universal, thermal, spectrum
can be understood from the observation that there is only one scale in
the problem, the Schwarzschild radius $R_S$. Other than indicating the
absorption cross-section of the black hole, the radiation can carry
no information. That is, it has maximum entropy, and so must have a thermal
spectrum. The temperature corresponds to a wavelength of $8\pi^2 R_S$.
\begin{figure}[tb]
\centerline{
\psfig{figure=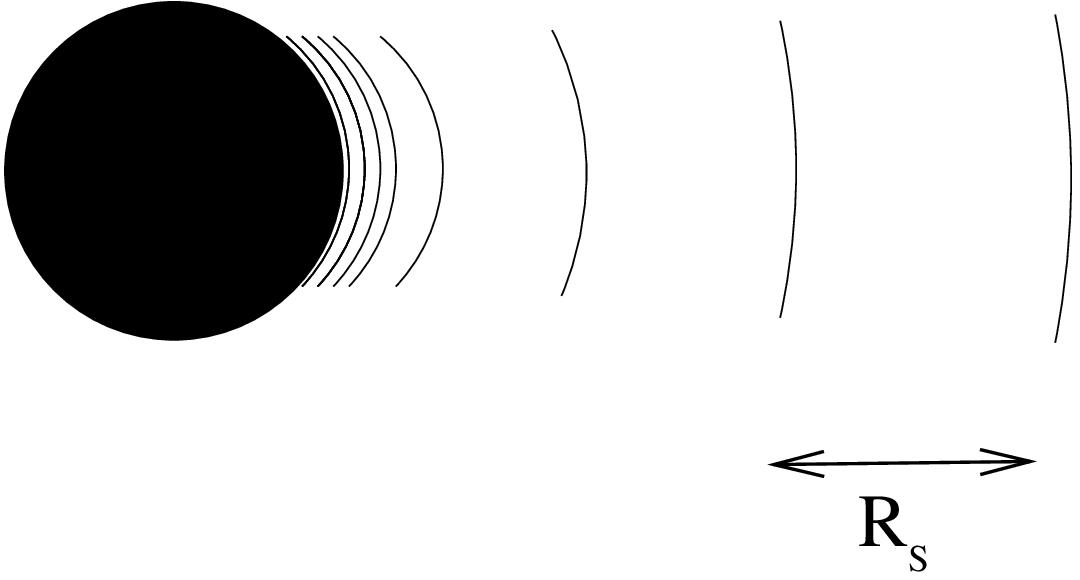,angle=-90,height=4cm}}
\fcaption{An outgoing mode typical of the Hawking radiation.
$T_H$ corresponds to a wavelength of $8\pi^2 R_S$.}
\label{hole}
\end{figure}
Since $\hbar$ is involved, one might think there is a second scale in the
problem, the Planck length
$L_P=(\hbar G/c^3)^{1/2}\sim 10^{-33}\, {\rm cm}$. However, $G$ does not enter
into Hawking's result because the gravitational interaction plays no role.
Only when one includes the back-reaction does $G$ or $L_P$ enter.
Indeed, the lifetime of the hole is of order $R_S^3/L_P^2c$.
(This is roughly $10^{54}$ times the present age of the universe
for a stellar mass black hole. For a primordial black hole of mass
$\sim10^{15}{\rm gm}$ however the lifetime is only the present age of the
universe.)

Upon further reflection however, it is not so clear that the Planck scale
physics is irrelevant to black hole
radiation\cite{Unruhexp,tHooft,Jacobson1,Jacobson2}.
Consider a vacuum fluctuation
mode that ends up at infinity populated with Hawking radiation. This mode
emerges from the vicinity of the horizon, and must climb out of a very deep
well (see Fig. \ref{hole}).
It is tremendously redshifted on the way out. If the mode has a
wavelength $\lambda_{out}$ when observed a time $t$ after the hole formed,
then its initial wavelength $\lambda_{in}$ must be of order
$\exp(-t/2R_s)\lambda_{out}$, which is very quickly way below the Planck
length. For instance if $\lambda_{out}\sim R_S$, then $\lambda_{in}\sim L_P$
when $t\sim 2R_S\ln(R_S/L_P)$, which is about a millisecond for a solar mass
black hole.
Hawking's original calculation assumed that vacuum fluctuations are
scale invariant down to such arbitrarily sub-Planckian dimensions.
If this assumption is no good, then the true behavior of quantum
fields around a black hole could be quite different.

The aim of these notes is both to review the standard
understanding of the Hawking effect, and to discuss the modifications
to this understanding that might be required by new physics at
short distances.
In particular, I first review the Unruh effect, namely, the fact that
an accelerated observer in Minkowski spacetime sees the vacuum as a thermal
state. This is then used to give an elementary explanation of the
origin of the Hawking effect, and to argue that, whatever short distance
field dynamics occurs on sub-Planckian scales, it will give rise to the
Hawking effect as long as it produces an ordinary vacuum structure
for the outgoing modes near the horizon at scales $\lambda$ satisfying
$L_P<\lambda\ll R_s$. The order of magnitude of deviations from the
thermal spectrum of Hawking radiation is estimated under various hypotheses.
In the remainder, the behavior of
a linear model with altered short distance physics---the Unruh
model---is discussed. This model consists of a scalar field
propagating on a black hole background with non-Lorentz-invariant higher
derivatives in the action. The higher derivatives lead to a nonlinear
dispersion relation at large wavevectors, but the field equations remain
linear.

In the following, unless otherwise indicated, units are adopted for which
$\hbar=c=G=k_B=1$.

\section{The Unruh effect}

A uniformly accelerated observer in flat spacetime
perceives the vacuum state of quantum fields
as a thermal state with
temperature $T_U=a/2\pi$, where $a$ is the
proper acceleration. This is called the Unruh effect\cite{Unruhnotes},
and $T_U$
is the Unruh temperature. To get a feeling for the magnitude of
this temperature, we can put $\hbar$ and $c$ back in:
\begin{equation}
T_U=\frac{\hbar}{c}\frac{a}{2\pi},
\end{equation}
so $T_U$ is roughly $\hbar$ divided by the time it takes for the
velocity to change by the speed of light.

The Unruh effect can be understood as a consequence of the
invariance of the vacuum under the symmetries of Minkowski
spacetime, translations and boosts, together with the spectral
condition that the energy of physical states be nonnegative.
I will give two derivations of this effect below, both of which
are valid for arbitrary interacting scalar fields in spacetime of any
dimension. (The generalization to fields of nonzero spin is essentially
trivial.) The second derivation, which uses a path integral method,
is much slicker and generalizes to the Hartle-Hawking state for
black holes and all other curved but static spacetimes with bifurcate Killing
horizons.
My reasons for giving the first derivation as well are threefold. First, it is
very concrete, second, it illustrates very nicely the dangers of
being sloppy with analyticity properties, and third, I have not seen it done in
the literature in this much generality in this simple manner.
A brief guide to the literature on these matters is provided in subsection
\ref{somerefs} below.

\begin{figure}[th]
\centerline{
\psfig{figure=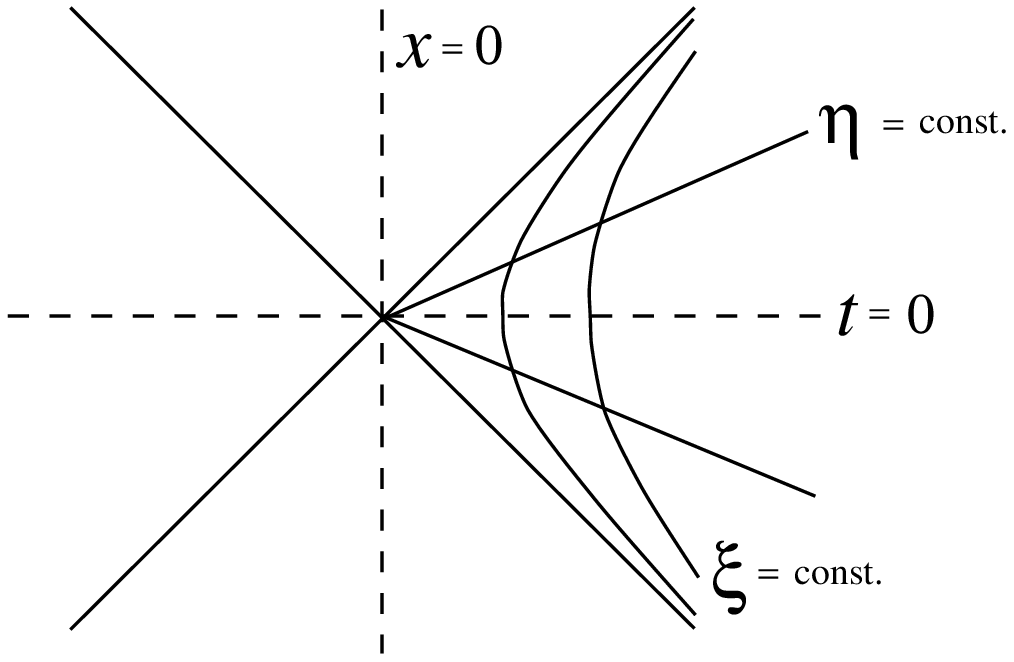,angle=-90,height=6cm}}
\fcaption{Two-dimensional flat spacetime in Minkowski and Rindler coordinates.
A hyperbola of constant $\xi$ is a uniformly accelerated timelike worldline
with proper acceleration $\xi^{-1}$. A boost shifts $\eta$ and  preserves
$\xi$.}
\label{Rindler}
\end{figure}

The Minkowski line element
in two dimensions can be written in both ``Cartesian" (Minkowski)
and ``polar" (Rindler) coordinates:
\begin{equation}
ds^2=dt^2-dz^2=\xi^2d\eta^2-d\xi^2
\label{mink}
\end{equation}
where the coordinates are related by
\begin{equation}
t=\xi\sinh\eta, \qquad
z=\xi\cosh\eta.
\label{tfn}
\end{equation}
The coordinates $(\eta,\xi)$ are nonsingular in the ranges
$\xi\in(0,\infty)$ and $\eta\in(-\infty,\infty)$, and cover the
``Rindler wedge" $x>|t|$ in Minkowski space (see Fig. \ref{Rindler}).
In the first form of the line element
the translation symmetries generated by
the Killing vectors
$\partial/\partial t$ and $\partial/\partial z$ are
manifest, and in the second form the boost symmetry
generated by the Killing vector
$\partial/\partial\eta$ is manifest. The latter is
clearly analogous to rotational symmetry in Euclidean space.
The full collection of translation and boost symmetries of Minkowski
spacetime is called the Poincar\'e group.

\subsection{Two-point function and KMS condition}
A thermal density matrix
$\rho=Z^{-1}\exp(-\beta H)$ has two identifying properties:
First, it is obviously {\it stationary}, since it commutes with the
Hamiltonian $H$. Second, because $\exp(-\beta H)$ coincides with
the evolution operator $\exp(-itH)$ for $t=-i\beta$, expectation
values in the state $\rho$ possess a certain symmetry under translation
by $-i\beta$ called the KMS condition\cite{Sewellbook,Haagbook}:
Let $\la A \ra_\beta$ denote the expectation value
$tr(\rho A)$, and let $A_t$ denote the time translation by $t$ of the operator
$A$. Using cyclicity of the trace we have
\begin{eqnarray}
\la A_{-i\beta}B\ra_\beta&=&
Z^{-1}tr\Bigl(e^{-\beta H} (e^{\beta H}Ae^{-\beta H})B\Bigr)\\
&=& Z^{-1}tr\Bigl(e^{-\beta H}BA\Bigr)\\
&=&\la BA\ra_\beta.
\label{KMS}
\end{eqnarray}
Note that for nice enough operators $A$ and $B$,
$\la A_{-i\tau}B\ra_\beta$ will be analytic in the strip $0<\tau<\beta$.
Now let us compare this behavior with that of the two-point function
along a uniformly accelerated worldline in the Minkowski vacuum.

If, as is usual, the vacuum state shares the symmetry of
Minkowski spacetime, then, in particular, the 2-point function
$G(x,x')=\la \phi(x)\phi(x')\ra$
must be a Poincar\'e invariant function of $x$ and $x'$. Thus it must
depend on them only through the invariant interval, so one has
$G(x,x')=f((x-x')^2)$ for some function $f$.
Now consider an ``observer" traveling along the hyperbolic trajectory
$\xi=a^{-1}$. This worldline has constant proper acceleration $a$,
and $a\eta$ is the proper time along the world line. Let us examine the
2-point function $G(\eta,\eta')\equiv G(x(\eta),x(\eta'))$ along this
hyperbola. It is clearly stationary, due to the boost invariance of the
vacuum, so it can only depend on $\eta$ and $\eta'$ through the
difference $\eta-\eta'$. Therefore
\begin{eqnarray}
G(\eta,\eta')&=&G\bigl(x(\eta-\eta'),x(0)\bigr)\\
                &=&f\bigl((x(\eta-\eta')-x(0))^2\bigr)\\
                &=&f\bigl(4a^{-2}\sinh^2[(\eta-\eta')/2]\bigr),
\end{eqnarray}
where the third equality follows from (\ref{tfn}). Now,
since $\sinh^2(\eta/2)$ is periodic under translations
of $\eta$ by $2\pi i$,
it appears that $G(\eta,\eta')$ is periodic under such translations in each
argument. In terms of the 2-point function the KMS condition implies
$G(\eta,\eta'+i\beta)=G(\eta',\eta)$, which is {\it not} the same
as translation invariance in each argument. Does this mean that
in fact the 2-point function in the Minkowski vacuum along the
accelerated worldline is {\it not} thermal? The answer is ``no", because
the above ``proof" that $G(\eta,\eta')$ is periodic was bogus.
First of all, a
Poincar\'e invariant function of $x$ and $x'$ need not depend only on
the invariant interval. It can also depend on the invariant step-function
$\theta(x^0-x'^0)\theta((x-x')^2)$. More generally, the analytic properties of
the
function $f$ have not been specified, so one cannot conclude from the
periodicity of $\sinh^2(\eta/2)$ that $f$ itself is periodic.
For example, $f$ might involve the square root, $\sinh(\eta/2)$, which is
{\it anti}-periodic. In fact, this is just what happens.

To reveal the analytic behavior of $G(x,x')$, it is necessary to
incorporate the condition that the spacetime momenta of states in the
Hilbert space lie inside or on the future light cone. One can show
(by inserting a complete set of states between the operators)
that this implies there exists an integral representation for the
2-point function of the form
\begin{equation}
G(x,x')=\int d^nk\, \theta(k^0) J(k^2) e^{-ik(x-x')},
\end{equation}
where $J(k^2)$ is a function of the invariant $k^2$ that vanishes
when $k$ is spacelike. Now let us evaluate $G(\eta,\eta')$
along the hyperbolic trajectory. Thanks to boost invariance, it suffices to
consider $\eta'=0$. Thus we have
\begin{equation}
G(\eta,0)=\int d^nk\, \theta(k^0) J(k^2)
e^{-ia^{-1}[k^0\sinh\eta -k^1(\cosh\eta-1)]}.
\end{equation}
Now it still looks as though this is periodic in translation of $\eta$ by
$i2\pi m$ for any integer $m$, but closer inspection reveals the difficulty
that the integral fails to be convergent if the imaginary part of $\sinh\eta$
becomes positive. One has
\begin{eqnarray}
\sinh(\eta-i\theta)&=&\sinh\eta\, \cos\theta - i \cosh\eta\, \sin\theta\\
\cosh(\eta-i\theta)&=&\cosh\eta\, \cos\theta - i \sinh\eta\, \sin\theta.
\end{eqnarray}
Since $k^0\ge|k^1|$ the integral is seen to converge for
$0<\theta<\pi$, but one cannot go past $\pi$. To see if the function has
an analytic extension beyond $\theta=\pi$ and, if so, to determine the
value at $\theta=2\pi$, one needs a different representation of $G(\eta)$.
A suitable representation can be obtained by observing that Lorentz invariance
allows us to
transform to the frame in which $x-x'$ has only a time component which is
given by the invariant norm $[(x-x')^2]^{1/2}=2a^{-1}\sinh(\eta/2)$.
Thus we have
\begin{equation}
G(\eta,0)=\int d^nk\, \theta(k^0) J(k^2)
e^{-i2a^{-1}k^0\sinh(\eta/2)}.
\end{equation}
In this form it is clear that $\eta-i\theta$ can be extended all the
way down to $\theta=2\pi$. Since $\sinh[(\eta-i2\pi)/2]=-\sinh(\eta/2)=
\sinh(-\eta/2)$, we can finally conclude that
$G(\eta-i2\pi,0)=G(-\eta,0)=G(0,\eta)$, which is the KMS condition (\ref{KMS}).

\subsection{The vacuum state as a thermal density matrix}
\label{BW}

The essence of the Unruh effect is the fact that the density matrix describing
the Minkowski vacuum, traced over the states in the region $z<0$, is
precisely a Gibbs state for the boost Hamiltonian $H_B$
at a ``temperature" $T=1/2\pi$:
\begin{equation}
Tr_{z<0}\, |0\ra\la0|=Z^{-1}\exp(-2\pi H_B),
\label{Gibbs}
\end{equation}
\begin{equation}
H_B=\int T_{ab}(\partial/\partial\eta)^a d\Sigma^b
\label{HB}
\end{equation}
This rather amazing fact has been proved in varying degrees of rigor
by many different authors.
A sloppy path integral argument making it
very plausible will be sketched below.

Since the boost Hamiltonian has dimensions of action rather than energy,
so does the ``temperature". To determine the local temperature seen by an
observer following a given orbit of the Killing field, note
from (\ref{mink}) that the norm
of the Killing field $\partial/\partial\eta$ on the orbit $\xi=a^{-1}$
is $a^{-1}$, whereas the observer has unit 4-velocity. If the Killing field
is scaled by $a$ so as to agree with the unit 4-velocity at
$\xi=a^{-1}$, then the boost Hamiltonian (\ref{HB}) and temperature are scaled
in the same way.
Thus the temperature appropriate to the observer at $\xi=a^{-1}$ is
$T=a/2\pi$. Since $a$ is the proper acceleration
of this observer, we recover the Unruh temperature defined above.
Alternatively, the two-point function defined by (\ref{Gibbs}) along the
hyperbola obviously satisfies the KMS condition
relative to boost time $\eta$ at temperature $1/2\pi$. When expressed
in terms of proper time $a\eta$, this corresponds to the temperature
$a/2\pi$.

One can view the relative coolness of the state at larger values of
$\xi$ as being due to a redshift effect---in this case a Doppler shift---
as follows.
Suppose a uniformly accelerated observer at $\xi_0$ sends some of the
thermal radiation he sees to another uniformly accelerated observer at
$\xi_1>\xi_0$. This radiation will suffer a redshift given
by the ratio of the norms of the Killing field: say $p$ is the
spacetime momentum
of the radiation. Then $p\cdot(\partial/\partial\eta)$ is
conserved\cite{Waldbook},
but the energy locally measured by the uniformly accelerated observer is
$ p\cdot(\partial/\partial\eta)/|\partial/\partial\eta|$, so that
$E_1/E_0=|\partial/\partial\eta|_0/|\partial/\partial\eta|_1$. This is
precisely the same as the ratio $T_1/T_0$ of the locally measured temperatures.
At infinity $|\partial/\partial\eta|=\xi$ diverges, so the temperature drops to
zero, which is consistent with the vanishing acceleration of the boost orbits
at infinity.

The path integral argument to establish (\ref{Gibbs}) goes like this:
Let $H$ be the Hamiltonian generating ordinary time translation in Minkowski
space. The vacuum $|0\ra$ is the lowest energy state, and we suppose it has
vanishing energy: $H|0\ra=0$. If $|\psi\ra$ is any state with nonzero
overlap with the vacuum, then $\exp(-\tau H)|\psi\ra$ becomes proportional
to $|0\ra$ as $\tau$ goes to infinity. That is, the vacuum wavefunctional
$\Psi_0[\phi]$ for a field $\phi$ is proportional to $\la\phi|\exp(-\tau
H)|\psi\ra$
as $\tau\rightarrow\infty$.
Now this is just a matrix element of the evolution operator between
imaginary times $\tau=-\infty$ and $\tau=0$,  and such matrix elements can be
expressed as a path integral in the ``lower half" of Euclidean space:
\begin{equation}
\Psi_0[\phi]=\int_{\phi(-\infty)}^{\phi(0)} {\cal D}\phi\, \exp(-I)
\label{PI}
\end{equation}
where $I$ is the Euclidean action.

\begin{figure}[bh]
\centerline{
\psfig{figure=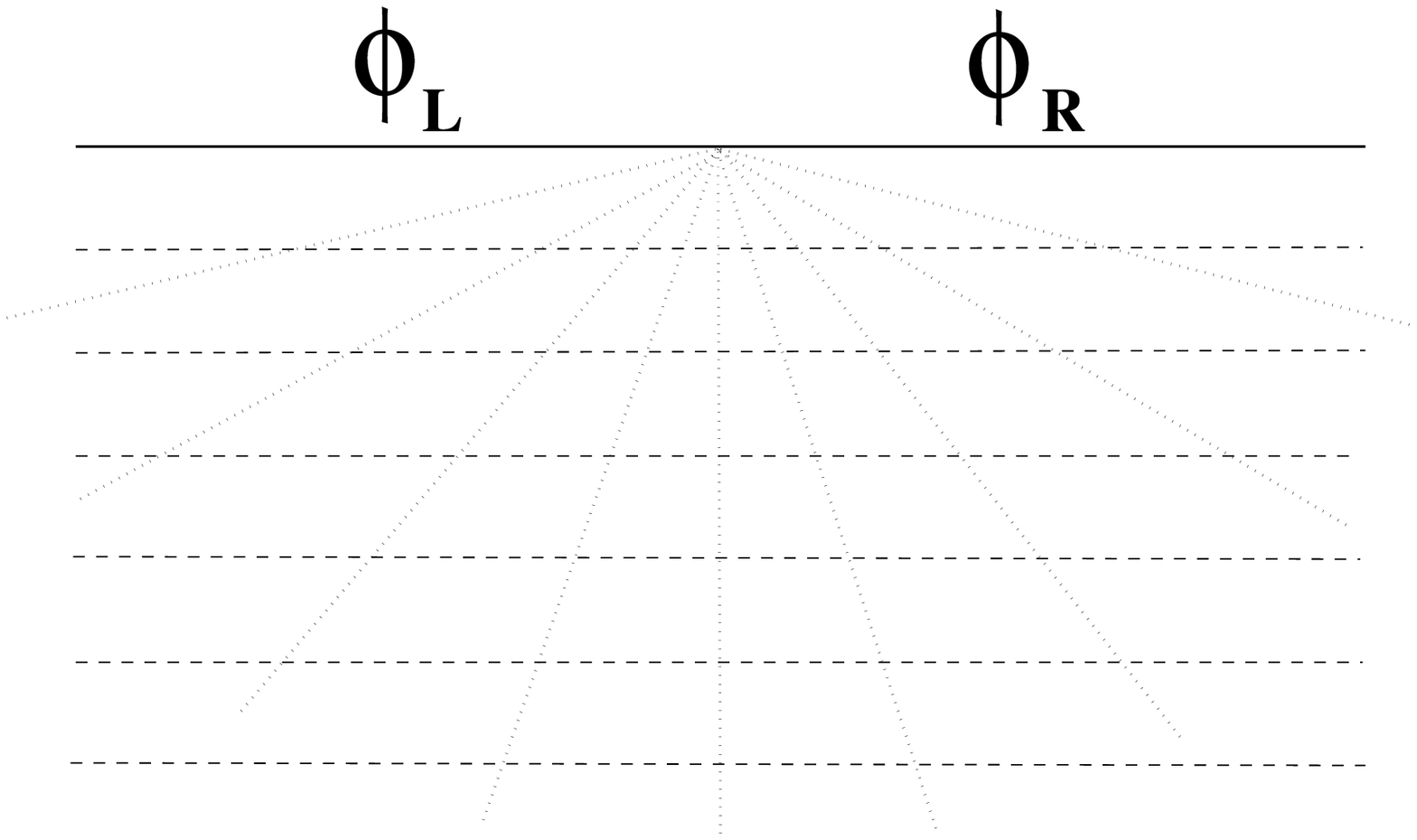,angle=-90,height=5cm}}
\fcaption{Time slicings of Euclideanized Minkowski space. The horizontal
lines are constant $\tau$ surfaces and the radial lines are constant
$\theta$ surfaces.}
\label{slicings}
\end{figure}

The key idea in recovering
(\ref{Gibbs}) is to look at (\ref{PI}) in terms of the angular
``time"-slicing of Euclidean space instead of the constant $\tau$ slicing. (See
 Fig. \ref{slicings}.)
The relevant Euclidean metric (restricted to two dimensions
for notational convenience) is given by
\begin{equation}
ds^2=d\tau^2 + d\sigma^2=\rho^2 d\theta^2 + d\rho^2.
\end{equation}
Adopting the angular slicing, the path integral (\ref{PI}) is seen to
yield an expression for the vacuum wavefunctional as a matrix element
of the boost Hamiltonian (\ref{HB}) which coincides with the generator of
rotations in Euclidean space:
\begin{equation}
\la\phi_L\phi_R|0\ra={\cal N}\la\phi_L|\exp(-\pi H_B)|\phi_R\ra,
\label{vac}
\end{equation}
where
$\phi_L$ and $\phi_R$ are the restrictions of the boundary value
$\phi(0)$ to the left and right half spaces respectively, and a
normalization factor ${\cal N}$ is included.
The Hilbert space ${\cal H}_R$ on which the
boost Hamiltonian acts consists of the field configurations on the right
half space $z>0$, and is being identified via reflection
(really, by reflection composed with CPT\cite{BisoWich,Sewell})
with the Hilbert space
${\cal H}_L$ of field configurations on the left half space $z<0$.
The entire Hilbert space is ${\cal H}={\cal H}_L\otimes{\cal H}_R$,
modulo the degrees of freedom at $z=0$. (The boundary conditions at $z=0$
are being completely glossed over here.) Now consider the vacuum
expectation value of an operator ${\cal O}_R$ that is localized on
the right half space:
\begin{eqnarray}
\la0|{\cal O}_R|0\ra
&=& \la 0|\phi_L\phi_R\ra\la\phi_L\phi_R|
{\cal O}_R|\phi_L'\phi_R'\ra\la\phi_L'\phi_R'|0\ra\\
&=& {\cal N}^2\la\phi_L|\exp(-\pi H_B)|\phi_R\ra
\la\phi_R|{\cal O}_R|\phi_R'\ra\la\phi_R'|\exp(-\pi H_B)|\phi_L\ra\\
&=& {\cal N}^2 tr\bigl(e^{-2\pi H_B}{\cal O}_R\bigr),
\label{expR}
\end{eqnarray}
where summation over intermediate states is implicit, and (\ref{vac})
was used in the second equality. This shows that, as far as observables
located on the right half space are concerned, the vacuum state is
given by the thermal density matrix (\ref{Gibbs}). More generally,
this holds for observables localized anywhere in the
Rindler wedge.

This path integral argument directly generalizes to all
static spacetimes with a bifurcate Killing horizon, such as the
Schwarzschild and deSitter spacetimes\cite{LaFlamme,Jacobsonhh}. In the
general setting, the state defined by the path integral cannot be
called ``the" vacuum, but it is a natural state that is invariant
under the static Killing symmetry of the background and is
nonsingular on the time slice where the boundary values of the field
are specified, including bifurcation surface.

\subsection{Some references}
\label{somerefs}
The fact that the Minkowski vacuum is a thermal state for the
boost Hamiltonian was proved in axiomatic quantum field theory
by Bisognano and Wichmann\cite{BisoWich}, as a theorem about the action
of complex Lorentz transformations on the vacuum. The relevance of this
theorem to the Unruh and Hawking effects was recognized by Sewell\cite{Sewell},
who generalized the framework to curved spacetimes. In completely
independent work (as far as I know) the path integral
argument has been given by many authors, perhaps the first being
Unruh and Weiss\cite{UnruhWeiss}. The review articles by Takagi\cite{Takagi}
and by Fulling and Ruijsenaars\cite{FullRuij} cover various aspects
of the relation between acceleration and temperature in quantum field theory,
and contain many other references.

\section{Hawking radiation}

The Hawking effect can be understood as a kind of ``gravitational
Unruh effect" as follows\cite{Unruhnotes,DeWitt}.
Consider a static, spherically symmetric,
asymptotically flat black hole spacetime. This spacetime has a time translation
Killing field $\chi$ which is a unit timelike vector at infinity, becomes
null on the horizon, and is spacelike within the horizon. An observer
at fixed radius is uniformly accelerated, with an acceleration that
vanishes at infinity and diverges as the horizon is approached.
What can be said about the state of the outgoing modes as viewed from
infinity in such a spacetime? It should be consistent with redshifting from the
state of the same modes viewed from a vantage point
closer in to the black hole, however that of course does not determine
the state.

In order to infer the existence of Hawking radiation one must
assume some boundary condition on the state of the quantum field.
In principle, this should just be the assumption that the initial state
was not too pathological and contained matter that would later collapse
and form a black hole. However, as discussed in the introduction,
to connect this initial condition to the final state one must evolve
initially arbitrarily short wavelength field modes through the collapse.
Instead let us first consider a boundary condition that does not involve
us in the physics of the trans-Planckian domain. Thus let us suppose that
the state just outside the horizon, after the black hole has formed, looks
like the Minkowski vacuum at very short, but longer than Planckian,
distances. If this is the case,
then a static observer hovering just outside the horizon
with a tremendous acceleration should experience the Unruh effect, and will
perceive the state to be thermal at some very high temperature.
This thermal state is then redshifted to infinity, as previously discussed
in the case of flat spacetime. However a crucial difference now arises:
the norm of the Killing field $\chi$ is finite at infinity, rather than
divergent. Thus the Unruh radiation survives its trip to infinity, where it
arrives with a non-zero temperature. Let us see how this works quantitatively.

For concreteness let us work with a Schwarzschild black hole, whose line
element is given in Schwarzschild coordinates by
\begin{equation}
ds^2=(1-2M/r)dt^2-(1-2M/r)^{-1}dr^2-r^2(d\theta^2+sin^2\theta d\phi^2).
\end{equation}
The static Killing field is $\chi=\partial/\partial t$, and its norm
is given by
$$N(r)=(\chi\cdot\chi)^{1/2}=(1-2M/r)^{1/2}.$$
$N$ is also called the ``lapse" function, since it determines how much
proper time $ds$ corresponds to a given coordinate time $dt$ via
$ds=Ndt$ at fixed $(r,\theta,\phi)$.
The acceleration of a worldline at constant radius and angles is
found to be
\begin{equation}
a=N^{-1}(1-N^2)^2\kappa,
\label{a}
\end{equation}
where $\kappa=1/4M$ is the ``surface gravity" of the hole. At the horizon the
acceleration diverges
and at infinity it vanishes. The Unruh temperature corresponding to
the acceleration is $a/2\pi$. As viewed from infinity this temperature
suffers a redshift given (as explained in subsection \ref{BW})
by the ratio $N(r)/N(\infty)=N(r)$. Thus the Unruh temperature for
an observer at $r$, as viewed
from infinity, is
\begin{equation}
T_{U,\infty}(r)=(1-N^2)^2\, T_H,
\label{TUi}
\end{equation}
where $T_H=\kappa/2\pi=1/8\pi M$ is the Hawking temperature. As $r$ approaches
the horizon the redshifted Unruh temperature approaches the Hawking
temperature, while as $r$ goes to infinity it vanishes.

Since the value of $T_{U,\infty}(r)$ in (\ref{TUi}) very much
depends on $r$, we appear to have an inconsistency. How can the state viewed
from infinity have many different temperatures? It cannot, of course.
The inconsistency arises only if we assume that at each radius the
static observer experiences the usual Unruh effect, but this assumption
requires that along each of these worldlines
the 2-point function of the quantum
field has the same form as it would have in Minkowski space. We have
no basis for making such an assumption, except in the limit of extremely
short distances where we expect all states to coincide. Since a higher
Unruh temperature arises from a shorter distance probe, this assumption
should get better and better as the static world line approaches the horizon.
In this limit, $T_{U,\infty}(r)$ approaches the Hawking temperature, so
it is the Hawking temperature that is selected by the Minkowskian boundary
condition on the quantum state at very short distances.

There are of course many other states of the quantum field. For example,
there is the ``Boulware vacuum", i.e., the Fock vacuum devoid of positive
Killing-frequency excitations. In this state the static observers
see no particles at any radius.
However, this state is not generally believed to be the
state that arises when a black hole forms in a collapse process.
In fact if one evolves the state of a free quantum field on the background
of a collapsing black hole, one sees that the final state is the
``Unruh vacuum". Near the horizon these two states are tremendously
different. Whereas the Unruh vacuum looks like the Minkowski vacuum
at very short distances, the Boulware vacuum does not. For example,
the renormalized expectation value of the stress-energy tensor
in the Boulware vacuum blows up as the horizon is approached.

\section{Imposing a cutoff}

Now imagine there is a short distance cutoff of some kind on the validity of
ordinary field theory, so one cannot impose the vacuum condition as above
in the limit of arbitrarily short distances. What effect would that
have on the Hawking effect? Of course the answer depends on what kind of
short distance physics is imagined.
Let us suppose first that, whatever the details at the cutoff, the vacuum
boundary condition is still satisfied to a fair degree at scales longer than
the cutoff $l_c$ but much shorter than the Schwarzschild radius.
That is, we assume the out vacuum at scales $\lambda$ satisfying
\begin{equation}
l_c<\lambda\ll R_s.
\label{scales}
\end{equation}
One would then
expect the Hawking effect to still occur, but with small deviations from the
usual Hawking spectrum. The presence of the cutoff presumably limits how close
to the horizon the accelerated observer can be assumed to experience the Unruh
effect. If the minimum allowed lapse is $N_{\rm min}$, then we might expect
from (\ref{TUi}) deviations at least of order $N_{\rm min}^2$ from the usual
Hawking flux\cite{Jacobson1}, since even if the observer at $N_{\rm min}$
experiences the usual Unruh effect (which is presumably not quite the case),
the redshifted temperature (\ref{TUi}) will not agree with $T_H$. The value of
$N_{\rm min}$ would depend on the nature of the cutoff.

\subsection{Invariant cutoff}

Without breaking local Lorentz invariance, one can imagine a cutoff
that limits the acceleration for which the usual Unruh effect takes place.
This might be the case if there is a minimum proper time that should be
considered along the worldline.
If the acceleration (\ref{a}) is required to be less than
$l_c^{-1}$, then $N_{\rm min}\sim \kappa l_c\simeq l_c/R_s$.
In this case the above argument
suggests deviations of order $(l_c/R_s)^2$ from the usual Hawking effect.
This might be relevant, for instance, in string theory.

\subsection{non-Lorentz-invariant cutoff}
\label{nli}

Lorentz invariance of the laws of physics has been confirmed to
some precision up to some finite boost factor $\gamma$ relative to the cosmic
rest frame. The boosts that play a role in the Hawking effect however
grow exponentially without bound in time as $\gamma\sim\exp(t/2R_s)$.
Normally physicists do not have confidence extrapolating infinitely
beyond their observations, but the case of Lorentz invariance seems to
be an exception. Allowing this exception seems to force us into assuming that
there are an infinite number of field modes in any finite spatial region, no
matter how small. And all of these modes play a role in the Hawking effect
as usually formulated. It is therefore worthwhile to ask what would happen
to the Hawking radiation in the presence of a non-Lorentz-invariant
cutoff\cite{Jacobson1,Jacobson2}.
This question will be considered in a detailed model later in these
notes. For now, let us just assume as above that the physics produces
the out vacuum at scales $\lambda$ satisfying (\ref{scales}) {\it in the
free-fall frame of the black hole}, that is, in the frame carried in
to the black hole by observers who fall freely from far away, starting
at rest.

Of course this definition of the preferred frame is ambiguous.
In the spherically symmetric case we can restrict to radial trajectories,
and consider only those trajectories that are always infalling, ignoring
those that pass through the collapsing matter and later fall back.
More generally, the fundamental cutoff theory would be needed to even formulate
the nature of the cutoff. But in order to explore the qualitative implications
of such a cutoff, it seems reasonable to stick to the spherically symmetric
case and to impose the cutoff in the falling frame defined above.

\begin{figure}[tb]
\centerline{
\psfig{figure=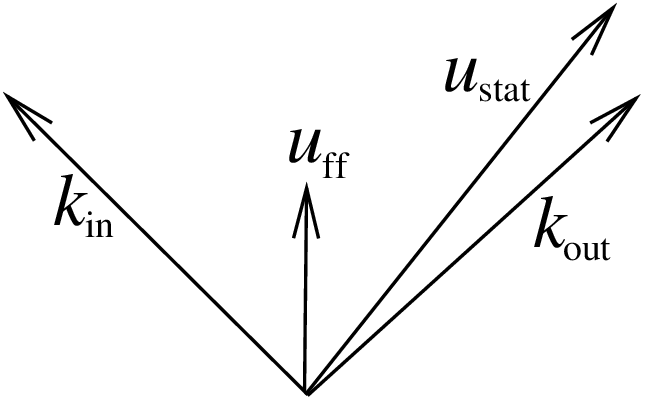,angle=-90,height=4cm}}
\fcaption{The null cutoff wave-vectors and the four-velocities of the
freely falling and static observers. The diagram is drawn in the free-fall
frame, so the static observer appears to be moving outward at high speed.}
\label{cutoff}
\end{figure}

The cutoff for the static observer is Doppler shifted relative to the cutoff
for the free-fall observer. The outgoing modes are redshifted while the
ingoing modes are blueshifted. Let $k_{c, {\rm in}}$ and $k_{c,{\rm out}}$
denote ingoing and outgoing null wave-vectors with the cutoff wavelength,
and let $u_{\rm ff}$ and $u_{\rm stat}$ denote the unit 4-velocities
of the free-fall and static observers (see Fig. \ref{cutoff}).
Then the cutoff frequency is
given in the free-fall frame by
$\o_c= k_{c, {\rm in}}\cdot u_{\rm ff}=k_{c,{\rm out}}\cdot u_{\rm ff}$,
while it is given in the static frame by
$\o_{c,{\rm in}}^{\rm stat}=k_{c, {\rm in}}\cdot u_{\rm stat}$ and
$\o_{c,{\rm out}}^{\rm stat}=k_{c, {\rm out}}\cdot u_{\rm stat}$
for the in- and out-going modes respectively.
Using these definitions near the horizon where $N\ll 1$ one finds
\begin{eqnarray}
l_{c,{\rm out}}^{\rm stat}&\simeq &2N^{-1}l_c\\
l_{c,{\rm in}}^{\rm stat}&\simeq &\textstyle{1\over2}Nl_c
\end{eqnarray}
For the Unruh effect we must require that the acceleration
timescale $a^{-1}$ is longer than the cutoff in the accelerated frame.
For the outgoing modes, which are relevant for the Hawking effect,
this means that $N/\kappa>N^{-1}l_c$. This non-Lorentz-invariant
cutoff thus leads to a much larger minimum lapse,
$N_{\rm min}\simeq(l_c/R_S)^{1/2}$. In this case one might expect deviations
from the Hawking flux of order $l_c/R_S$, which is still very small for
all but Planckian holes.

A quantum field theoretic calculation of the out state at infinity
starting from the out vacuum near the horizon at scales (\ref{scales})
in the free-fall frame
has also been carried out\cite{Jacobson2}. To do this calculation it is
necessary to work with localized wavepackets. Being careful about the
errors caused by the inevitable ``tails" of the wavepackets, an upper bound
of order $(l_P/R_S)^{-1/2}$ for the deviations from the Hawking spectrum
was found, which is much larger than the estimate obtained from the
Unruh effect argument above. However, this was only an upper bound.

\subsection{Horizon fluctuations}

So far I have been speaking of a cutoff in terms of a modification
of the field theory on a fixed background. On the other hand,
taking into account the back reaction and quantum fluctuations
in geometry one expects the horizon to fluctuate in some sense.
It is interesting to try to estimate the size of these fluctuations
and their possible effect on the Hawking radiation.
Let us consider the fluctuations in the horizon that would be
expected if the horizon is viewed as a system of Planck areas
fluctuating about an equilibrium configuration. In equilibrium the
typical entropy fluctuation is given by $\d S\sim 1$. Since
$S=A/4$ (where $A$ is the horizon area), we therefore expect
$\d A\sim1$. Now if we
assume the horizon consists of $A$ independent Planck areas
with random area fluctuation $\d a\sim R\d r/A\sim \d r/R$
(in four-dimensional spacetime),
then by the law of large numbers we have $\d A\sim A^{1/2}\d a\sim\d r$, so
$\d r\sim 1=l_P$.
This reasoning suggests that the horizon is fuzzy on scales of order
$l_P$ in the radial coordinate. If this is interpreted to mean that one should
not apply the acceleration temperature argument any closer to the horizon
than $r=R_S+l_P$, one finds for the minimum lapse
$N_{\rm min}\sim (l_P/R_S)^{1/2}$. This is the same result as one would
have obtained by imposing a cutoff at $l_P$ in the free-fall frame as
above.

\section{Models with a non-Lorentz-invariant cutoff}

In the previous section we argued crudely that the Hawking effect will
occur independently of short distance physics as long as the
out vacuum boundary condition holds near the horizon
for wavelengths satisfying
$l_c<\lambda\ll R_s$, up to small corrections for large black holes.
This leaves us with the following questions. Given a particular
theory with some new physics at short distances,
\begin{enumerate}
\item Does the abovementioned out vacuum boundary condition hold?
\item Exactly how large are the deviations from the the thermal Hawking flux?
\item Are the deviations from the thermal Hawking flux small even at
very short wavelengths?
\item Do the deviations for short wavelengths accumulate to make
a large difference in any physical quantity, such as the energy flux
or energy density?
\end{enumerate}

A class of models in which these questions can be quantitatively
addressed with relatively simple methods is obtained if the usual
wave equation for the quantum
field is modified by the addition of higher derivative terms, while
linearity of the field equation is preserved. Such a modification
can be covariant, or it can be done in a manner that violates
local lorentz invariance. I know of no suitable covariant
model for ordinary fields, since covariant higher derivatives
always lead to negative norm ghosts. However,
string theories can be Lorentz-covariant and ghost-free,
and it might be feasible and
interesting to study Hawking radiation in non-interacting string
theory. Attempts in this direction have been made but so far none
have been able to compute the spectrum of black hole radiation.
On the other hand, several {\it non}-covariant models ones have
been studied recently.
The mother of them all was Unruh's sonic black hole.
Although the subsequent models can all be interpreted in terms
of ordinary black holes without reference to fluid flow, I will
first describe the sonic black hole model, since it is very
interesting and provides good motivation for the other models.

\subsection{Sonic black hole}
Unruh\cite{Unruhexp} pointed out in 1981 that perturbations of an
irrotational, inviscid, barotropic ($\rho=\rho(p)$) fluid flow behave like
a massless field propagating in a curved spacetime whose metric is
determined by the background flow\cite{}. If the flow is supersonic somewhere,
then there is a horizon in the effective spacetime geometry, and
one has a model of a black hole. Quantizing the fluid perturbations,
Unruh argued that, provided the out vacuum boundary condition holds
just outside the sonic horizon, the sonic black hole will radiate
thermal phonons at the temperature $\kappa_s/2\pi$, where
$\kappa_s$ is the gradient of the velocity field at the horizon.
(Numerically,  $\kappa_s/2\pi \simeq 10^{-7}{\rm K}\,
({\partial v\over \partial r}/{100{\rm m/s}\over 1{\rm mm}})$.)
This is already interesting, but the real reason Unruh invented this
model was in the hope of studying both the consequences of the molecular
nature of the fluid and the quantum backreaction. Since the molecular
structure has a preferred local frame---the comoving frame of the fluid---
the {\it molecular} fluid model lacks covariance in the fluid metric.
That is, the molecular fluid model displays a non-Lorentz-invariant cutoff.

Consider the sonic black hole with liquid Helium-4 as the
fluid\cite{Jacobson1}.
The sound field describes the quasiparticle excitations, and
satisfies the usual wave equation for long wavelengths. The atomic nature
of the fluid begins to show itself in the nonlinearity of the quasiparticle
dispersion relation $\omega(k)$ at shorter wavelengths. This nonlinearity
implies that the group velocity is $k$--dependent, at first dropping as $k$
grows. This has important
implications for the propagation of wavepackets in the sonic black hole
background\cite{}. In particular, one can infer
that an outgoing wavepacket propagated backwards in time toward the
sonic horizon will be blueshifted, hence its comoving group velocity
will drop. The packet will not quite make it all the way to the horizon, but
will stop where its comoving group velocity is equal to the negative of the
background flow velocity. This is an unstable situation, and what happens at
this stage is remarkable\cite{Unruhpc}: the packet is blueshifted a bit more,
the comoving group velocity drops a bit more, and the packet is swept back away
from the horizon by the (time-reversed) flow! This is {\it very} different from
the behavior of a wavepacket satisfying the ordinary wave equation, which
is to get squeezed more and more up against the black hole horizon, while
exponentially blueshifting.

\subsection{Unruh model}
It is not necessary to deal with liquid Helium to study the behavior
just described. All that is essential is the curvature of the dispersion
relation, and this can be modeled in linear field theory with higher
spatial derivative terms\cite{Unruhdumb}.
Since in the case of a fluid the dispersion
relation is physically specified in the comoving frame,
the comoving time derivative should be left unchanged.
Unruh studied the Hawking process in a model of this sort,
by numerically integrating the partial differential equation governing
the propagation of a wavepacket\cite{}.
Because of the reversal of wavepacket motion at the horizon,
it is possible to deduce the final state from an initial condition
on ingoing modes far from the horizon, even in a stationary background
(see Section \ref{puzzle} for further discussion of this issue).
Unruh found, to numerical accuracy
of his computation, that the usual Hawking flux is emitted from the
hole. Subsequently the same and similar models were studied using
using both analytical approximations\cite{BMPS}, and by first using
stationarity
to reduce the problem to the solution of ordinary
differential equations\cite{CorlJaco}. Using the last described method, it
was possible to reliably compute deviations from the usual thermal
spectrum of Hawking radiation.

\begin{figure}[bt]
\centerline{
\psfig{figure=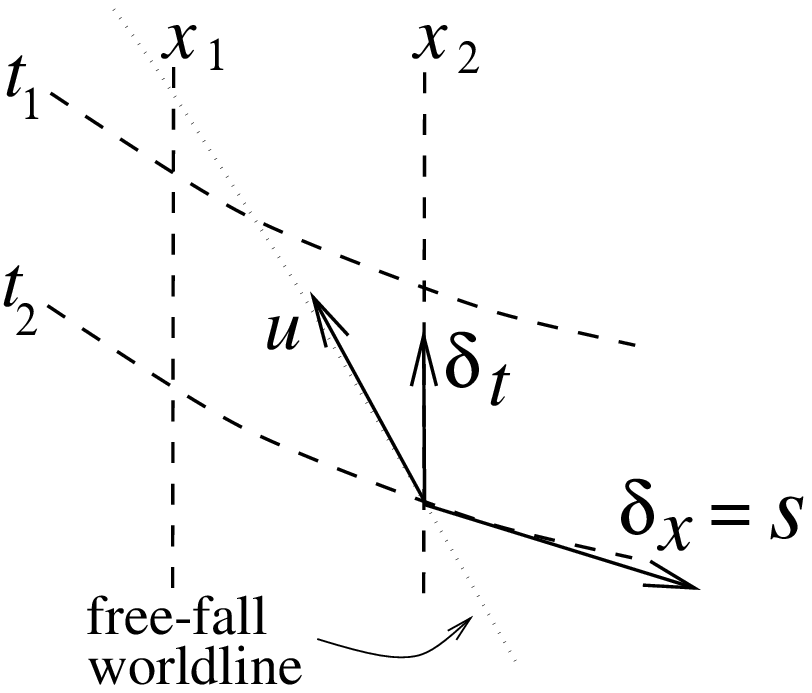,angle=-90,height=6cm}}
\fcaption{A patch of spacetime showing a free-fall trajectory
and some $t$ and $x$ (Lema\^{\i}tre-like) coordinate lines. $u$ and $s$ are
orthonormal vectors, and the derivative along $s$ is modified,
while that along $u$ is just the partial derivative. The notations $\delta_t$
and $\delta_x$ denote $\partial/\partial t$
and $\partial/\partial x$ respectively, and $\delta_t$ is the Killing vector.}
\label{lemaitre}
\end{figure}

Here I will describe the Unruh model and some of its generalizations
without further reference to the fluid flow analogy.
The model consists of a free, hermitian scalar field propagating in a two
dimensional black hole spacetime. The dispersion relation for the field
lacks Lorentz invariance, and is specified in the free fall frame of the
black hole, that is, the frame carried in from the rest frame at infinity by
freely falling trajectories.
Let $u^\a$ denote the unit vector field tangent to
the infalling worldlines, and let $s^\a$ denote the orthogonal,
outward pointing, unit vector, so that $g^{\a\b}=u^\a u^\b-s^\a s^\b$
(see Fig. \ref{lemaitre}).
The action is assumed to have the form:
\begin{equation}
S={1\over2}\int d^2x \, \sqrt{-g} g^{\a\b} {\cal D}_\a\phi^* {\cal D}_\b\phi,
\label{action}
\end{equation}
where the modified differential operator ${\cal D}_\a$ is defined by
\begin{eqnarray}
u^\a {\cal D}_\a &=& u^\a \partial_\a\\
s^\a {\cal D}_\a &= &\hat{F}(s^\a \partial_\a).
\end{eqnarray}
The time derivatives in the local free fall frame are thus left unchanged,
but the orthogonal spatial derivatives are replaced by
$\hat{F}(s^\a \partial_\a)$.
The function $\hat{F}$ determines the dispersion relation. For the moment it
will be left unspecified.
Invariance of the action (\ref{action}) under
constant phase transformations of $\phi$ guarantees that there is a conserved
current for solutions and a conserved ``inner product" for pairs of solutions
to the equations of motion. However, since ${\cal D}_\a$ is {\it not} in
general a derivation, simple integration by parts is not allowed in obtaining
the equations of motion or the form of the current. We shall obtain these below
after further specifying the model.

The black hole line elements we shall consider are static and have the form
\begin{equation}
ds^2=dt^2-(dx-v(x)\, dt)^2.
\end{equation}
This is a generalization of the Lema\^{\i}tre line element for the
Schwarzschild
spacetime, which is given by $v(x)=-\sqrt{2M/x}$ (together with the usual
angular part). We shall assume $v<0$, $dv/dx>0$, and
$v\rightarrow v_o$ as $x\rightarrow\infty$.
$\partial_t$ is a Killing vector, of squared norm
$1-v^2$, and the event horizon is located at $v=-1$.
The curves given by $dx-v\, dt=0$ are timelike free fall worldlines
which are at rest (tangent to the Killing vector) where $v=0$.
Since we assume $v<0$ these are {\it ingoing} trajectories.
$v$ is their coordinate velocity, $t$ measures proper time along them,
and they are everywhere orthogonal to the constant $t$ surfaces.
(See Fig. \ref{lemaitre}.)
We shall refer to the function $v(x)$ as the {\it free-fall velocity}.
The asymptotically flat region corresponds to $x\rightarrow\infty$.

In terms of the notation above, the orthonormal basis vectors adapted to the
free fall frame
are given by $u=\partial_t+v\partial_x$ and $s=\partial_x$, and
and in these coordinates $g=-1$. Thus the action (\ref{action}) becomes
\begin{equation}
S={1\over2}\int dtdx\,
\Bigl(|(\partial_t+v\partial_x)\phi|^2 -|\hat{F}(\partial_x)\phi|^2\Bigr).
\end{equation}
If we further specify that $\hat{F}(\partial_x)$ is an odd function of
$\partial_x$,
then integration by parts yields the field equation
\begin{equation}
(\partial_t+\partial_x
v)(\partial_t+v\partial_x)\phi=\hat{F}^2(\partial_x)\phi.
\label{eom}
\end{equation}
The conserved inner product in this case is given by
\begin{equation}
 (\phi,\psi) = i \int dx\,  \Bigl(\phi^{*}
(\partial_{t} + v\partial_{x})\psi - \psi(\partial_{t} + v\partial_{x})
\phi^{*}\Bigr),
\label{inner}
\end{equation}
where the integral is over a constant $t$ slice and is independent of $t$  if
$\phi$, $\psi$ satisfy the field equation (\ref{eom}). The inner product can of
course be
evaluated on other slices as well, but it does not take the same simple form
on other slices.

\begin{figure}[h]
\centerline{
\psfig{figure=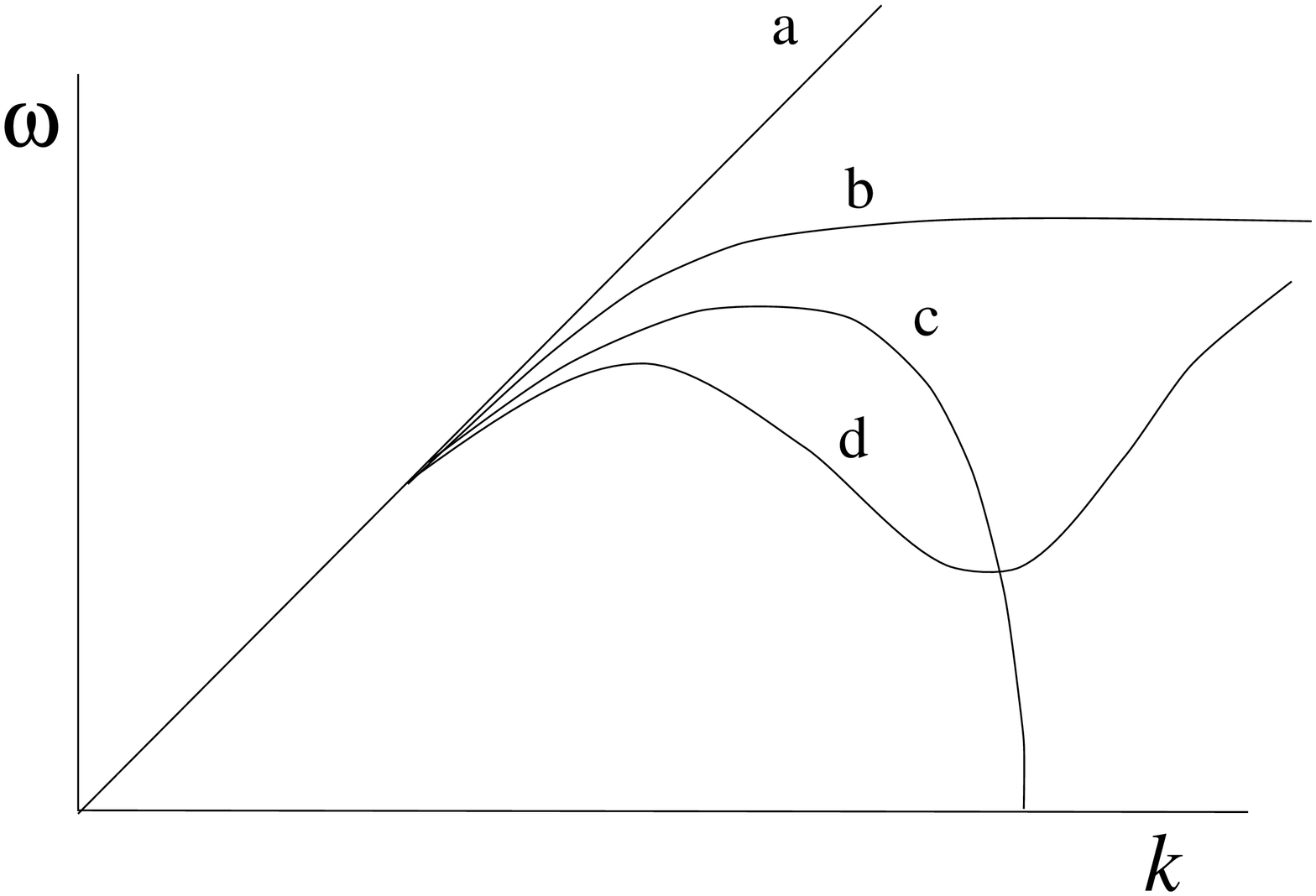,angle=-90,height=5cm}}
\fcaption{The dispersion relations for (a) the massless wave equation,
(b) the Unruh model (\ref{FUN}), (c) the quartic model (\ref{F}),
and (d) liquid helium-4.}
\label{disps}
\end{figure}

The dispersion relation for this model in flat spacetime, or in the local
free fall frame, is given by
\begin{equation}
\o^2=F^2(k),
\label{disp}
\end{equation}
where $F(k)\equiv-i\hat{F}(ik)$.
Unruh's choice for the function $F(k)$ has the property that
 $F^{2}(k)$ = $k^{2}$ for $k \ll k_{0}$ and $F^{2}(k)$ = $k_{0}^2$ for $k \gg
k_{0}$, where $k_{0}$ is a wavevector characterizing the scale of the new
physics. We usually think of $k_0$ as being around the Planck mass.
Specifically, he considered the functions
\begin{equation}
F(k)=k_0\{\tanh[(k/k_0)^n]\}^{1\over n}.
\label{FUN}
\end{equation}
 Of course there are many other modifications one could consider.  Perhaps the
simplest is given by
\begin{equation}
F^{2}(k) = k^{2}- k^4/k_{0}^{2}.
\label{F}
\end{equation}
This dispersion relation has the same behaviour
for small $k$ as the one above, but behaves quite differently for large $k$.
It is the one which was studied in Ref. \cite{CorlJaco}. These two dispersion
relations are plotted in Fig. \ref{disps} along with the dispersion
relations for the ordinary wave equation and for elementary excitations of
liquid helium-4.

\subsubsection{Quantization}
To quantize the field we assume the field operator $\hat\phi(x)$ is
self-adjoint
and satisfies the equation of motion (\ref{eom}) and the canonical commutation
relations. In setting up the canonical formalism, it is simplest to use the
time function and evolution vector for which only first order time
derivatives appear in the action. (Otherwise one must introduce extra momenta
which are constrained, and then pass to the reduced phase space.)
This just means that we define the momenta by
\begin{equation}
\pi=\delta L/\delta (\partial_t\phi)=(\partial_t +v\partial_x)\phi,
\label{pi}
\end{equation}
i.e., $\pi$ is the time derivative along the free-fall world lines.
The equal time canonical commutation relations are then
$[\phi(x),\pi(y)]=i\delta(x,y)$, as usual.

We define an annihilation operator corresponding to an initial data set
$f$ on a surface $\Sigma$ by
\begin{equation}
a(f)=( f,\hat{\phi}),
\label{C2}
\end{equation}
where the inner product is evaluated on $\Sigma$.
If the data $f$ is extended to a solution of the field
equation then we can evaluate the inner product in (\ref{C2}) on
whichever surface we wish. The hermitian adjoint of $a(f)$
is called the creation operator for $f$ and it is given by
\begin{equation}
a^{\dagger}(f)=-( f^*,\hat{\phi}).
\end{equation}
The commutation relations between these operators follow
from the canonical commutation relations satisfied by the
field operator. The latter are equivalent to
\begin{equation}
 [a(f),a^{\dagger}(g)]=( f, g),
\label{C3}
\end{equation}
provided this holds for all choices of $f$ and $g$.
Now it is clear that only if $f$ has positive, unit norm
are the appelations ``annihilation" and ``creation"
appropriate for these operators.
{}From (\ref{C3}) and the definition of the inner product
it follows identically that we also have the commutation
relations
\begin{equation}
[a(f),a(g)]=-( f, g^*),\qquad
[a^{\dagger}(f),a^{\dagger}(g)]=-( f^*, g).
\end{equation}

A Hilbert space of ``one-particle states" can be defined by
choosing a decomposition of the space $S$ of complex
initial data sets (or solutions to the field equation) into a
direct sum of the form $S=S_p\oplus S_p{}^*$, where all the
data sets in $S_p$ have positive norm and the space $S_p$
is orthogonal to its conjugate $S_p{}^*$. Then all of the
annihilation operators for elements of $S_p$ commute with
each other, as do the creation operators. A ``vacuum" state
$|\Psi\ra$ corresponding to $S_p$ is defined by the condition
$a(f)|\Psi\ra=0$ for all $f$ in $S_p$, and a Fock space of
multiparticle states is built up by repeated application
of the creation operators to $|\Psi\ra$.

It is not necessary to construct a specific Fock space in order to
study the physics of this system. In fact, any individual positive
norm solution $p$ defines annihilation and creation operators and
a number operator $N(p)=a^\dagger(p)a(p)$. The physical significance
of the number operator depends of course on the nature of $p$.

There are two types of positive norm wavepackets in which we are
interested. The first are those corresponding to the quanta of Hawking
radiation. These have positive Killing frequency, that is,
they are sums of solutions satisfying $\partial_t\phi=-i\o\phi$
with $\o>0$. It is not obvious that such solutions have positive norm
in the inner product (\ref{inner}), and in fact they do not in general.
However, using the fact that
the Killing frequency is conserved, we know that if a positive Killing
frequency wavepacket were to propagate out to infinity (or any other region
where $v=0$),
the integral for its norm would be manifestly positive. Since the
norm is conserved, this suffices.

The other type of positive norm wavepackets we shall employ are those
which correspond to particles as defined by the free-fall observers.
These have positive free-fall frequency, that is, they are sums of solutions
satisfying $(\partial_t+v\partial_x)\phi=-i\o'\phi$, with $\o'>0$,
on some time slice. These have manifestly positive norm, although the free-fall
frequency is {\it not} conserved.

\subsubsection{Computing the particle creation rate}
In this subsection we obtain the explicit expression for the particle creation
rate in terms of the norm of the negative frequency part of the ingoing
wavepacket that corresponds to a given outgoing wavepacket. Let
$\psi_{{\rm out}}$ denote a wavepacket solution of the field equation
(\ref{eom})
which at late times is outgoing and localized in the constant velocity region,
where it has only positive free-fall frequency components, and
only positive Killing frequency components.
Propagating this data backwards in time, a part will reflect off the
geometry outside the black hole and arrive back in the constant
velocity region again as a wavepacket $\psi_r$ with
wavevector components at the small negative $k(\omega)$ root
with positive free-fall frequency. (If the ordinary two-dimensional
wave equation were satisfied, this reflected part would vanish due to
conformal invariance. However the field equation (\ref{eom}) lacks conformal
invariance at high wavevectors, so in general there will be some reflection,
which will be extremely small unless the metric has short wavelength features.)
The rest of the wavepacket will reach within a distance of order $k_0^{-1}$
from the horizon, where it will undergo mode conversion and head back out,
ending up in the constant velocity region as a combination of positive
and negative free fall frequency wavepackets $\psi_{+}$ and $\psi_{-}$.
The mode conversion process will be explained in the next section.

Since the inner product is time independent we have
\begin{equation}
(\psi_{{\rm out}},\hat{\phi}) =  (\psi_{r},\hat{\phi})+
(\psi_{+},\hat{\phi}) + (\psi_{-},\hat{\phi}),
\end{equation}
or in terms of annihilation and creation operators,
\begin{equation}
a(\psi_{{\rm out}}) = a(\psi_{r})+ a(\psi_{+}) - a^{\dagger}(\psi_{-}^*).
\end{equation}
We assume that the state of the field at early times is the free-fall vacuum,
$|{\rm ff}\rangle$, which satisfies
$a(\psi)|{\rm ff}\rangle$ = 0 for any ingoing positive free-fall
frequency wavepacket $\psi$. The particle creation in the packet
$\psi_{{\rm out}}$, characterized by the expectation value of the number
operator, is thus given by the norm of $\psi_{-}$:
\begin{equation}
N(\psi_{\rm out})= \langle{\rm ff}|
a^{\dagger}(\psi_{{\rm out}})a(\psi_{{\rm out}})
|{\rm ff}\rangle =
-(\psi_{-},\psi_{-}).
\label{number}
\end{equation}

\section{The Hawking effect and mode conversion}

In this section I will first describe the results of the particle production
calculation in the Unruh model. These results will then be explained with the
help of the WKB approximation.
The resulting picture has two essential features: reversal of group velocity
without reflection, and ``mode conversion" from one branch of the dispersion
relation to another. Interestingly, both these phenomena can occur for linear
waves
in inhomogeneous plasmas\cite{Skiff,Stix,Swanson}, and undoubtedly occur in
many other settings as well. At the turn-around point the pure
WKB approximation breaks down, and partial mode conversion from a positive
free-fall frequency to a negative free-fall frequency wave takes place.
This mode conversion gives rise to the Hawking effect.

\begin{figure}[p]
\centerline{\psfig{figure=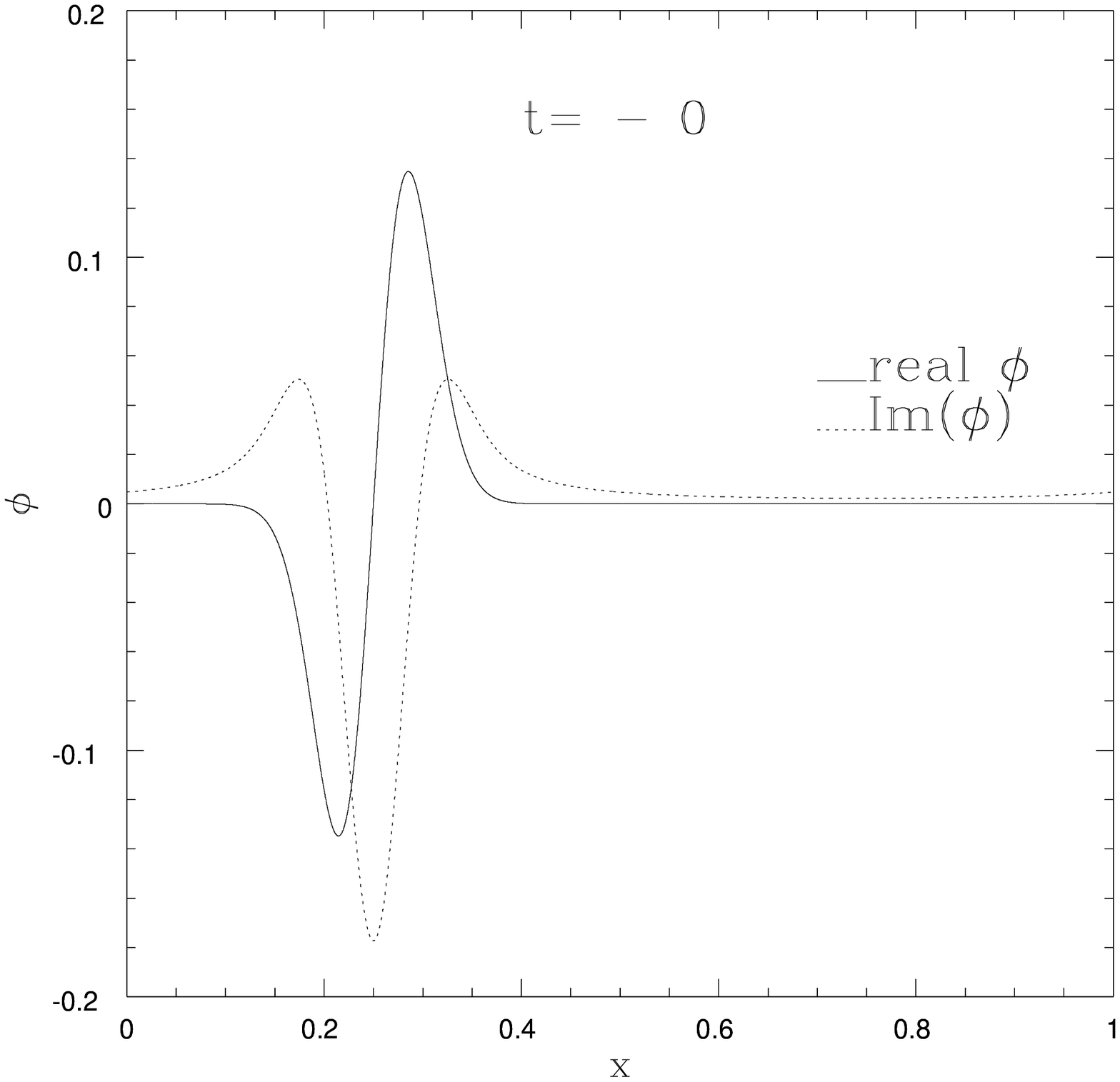,angle=0,height=6.5cm}\hfill
\psfig{figure=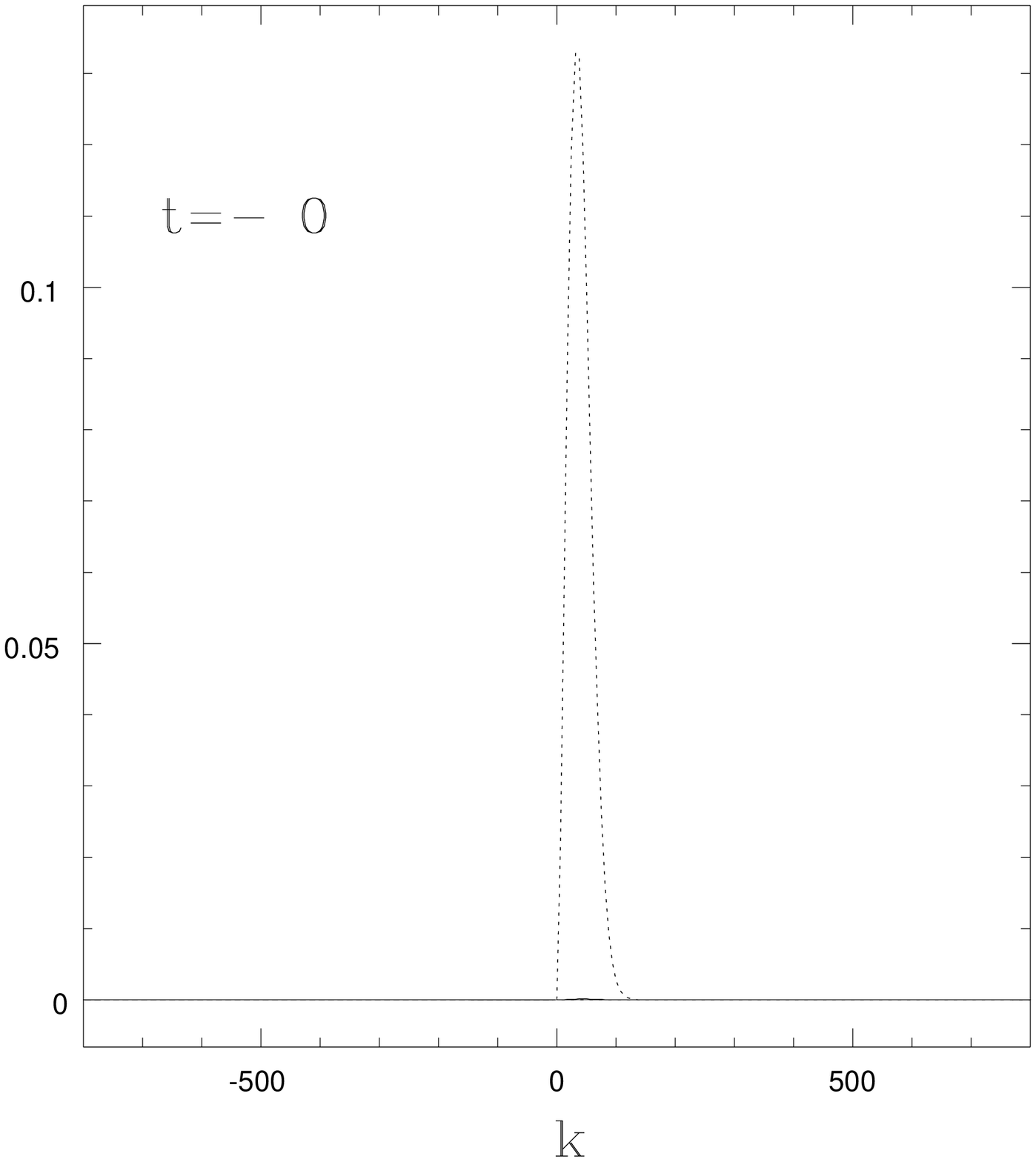,angle=0,height=6.5cm}}
\centerline{\psfig{figure=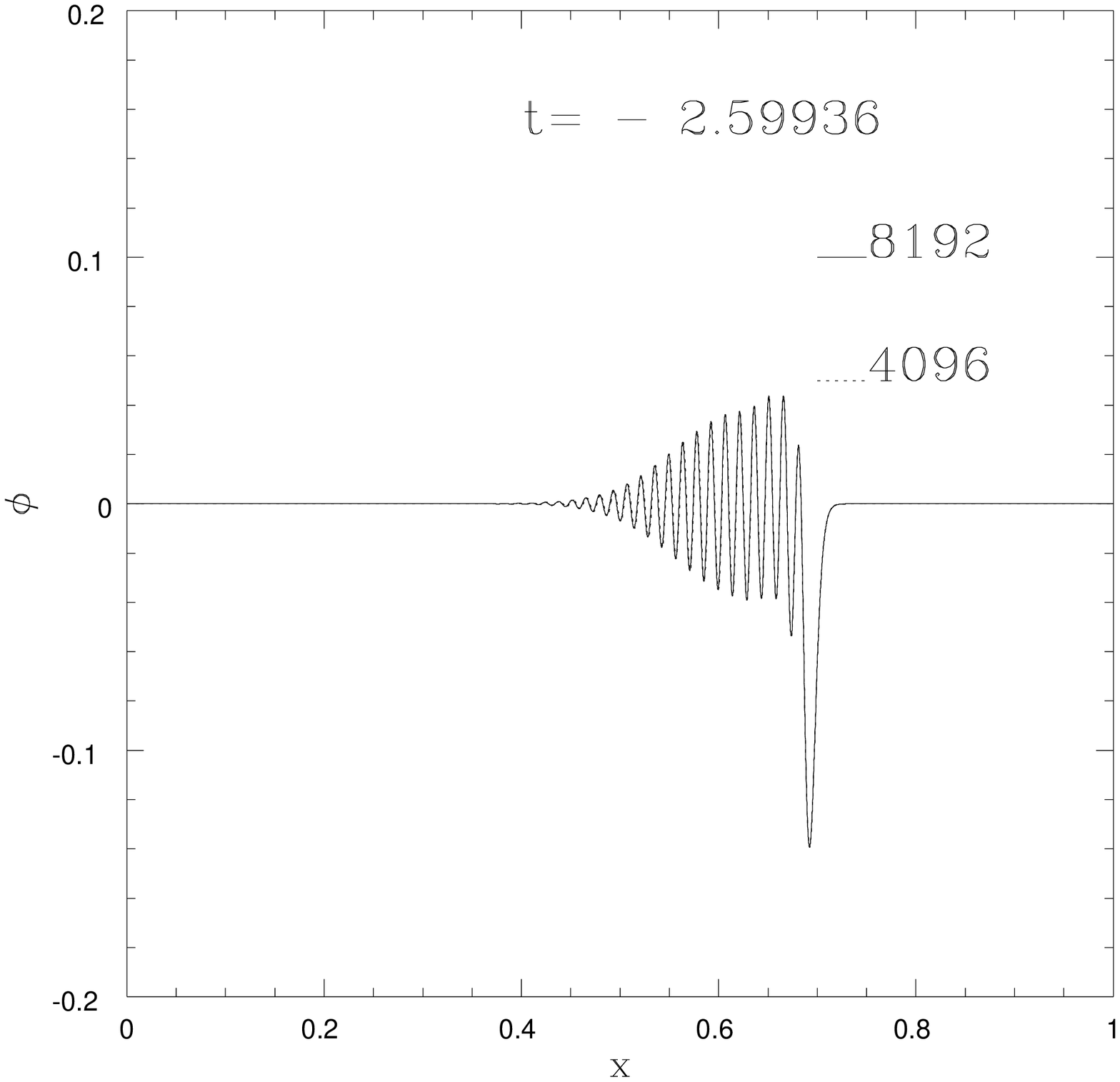,angle=0,height=6.5cm}\hfill}
\centerline{\psfig{figure=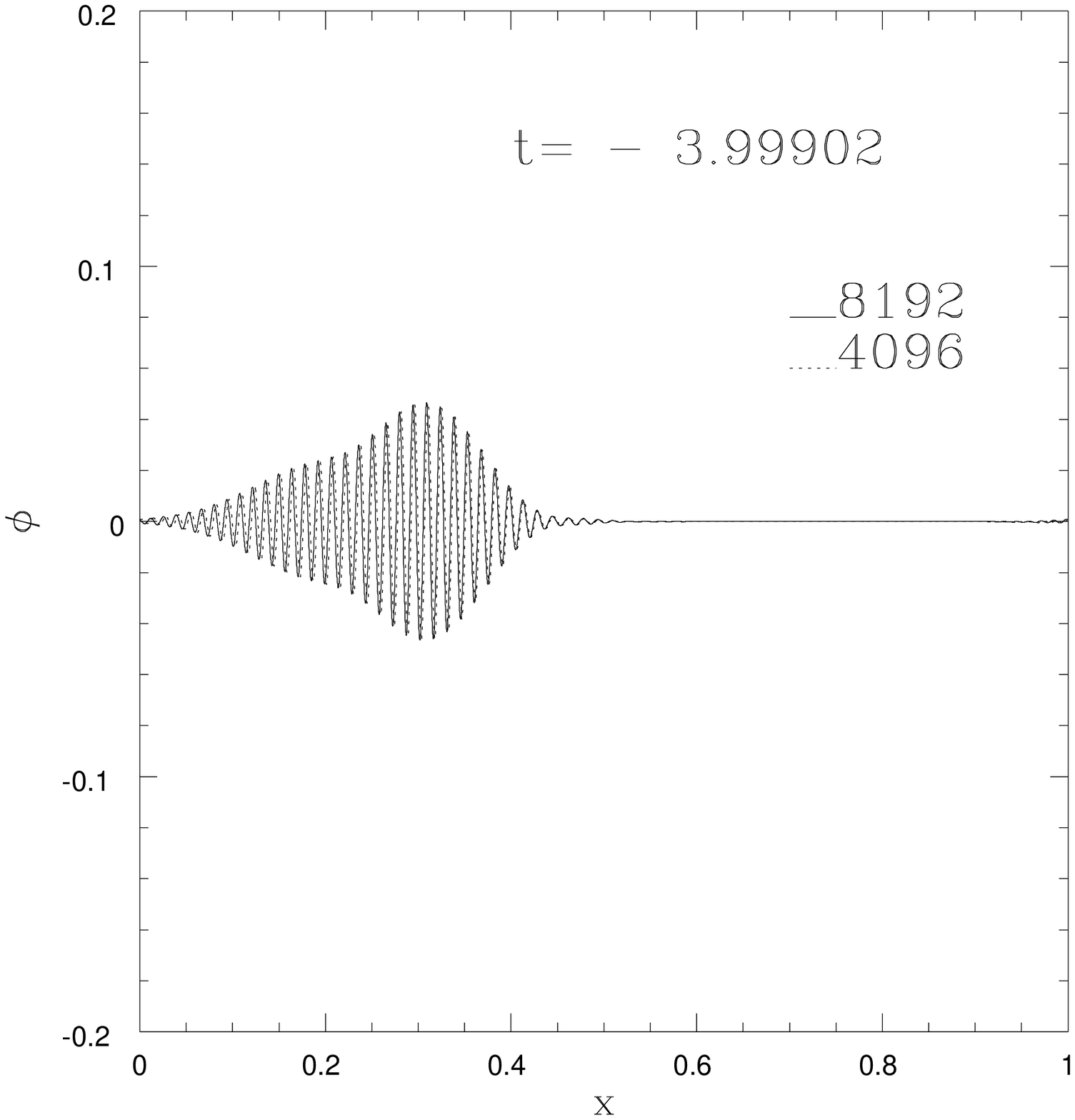,angle=0,height=6.5cm}\hfill
\psfig{figure=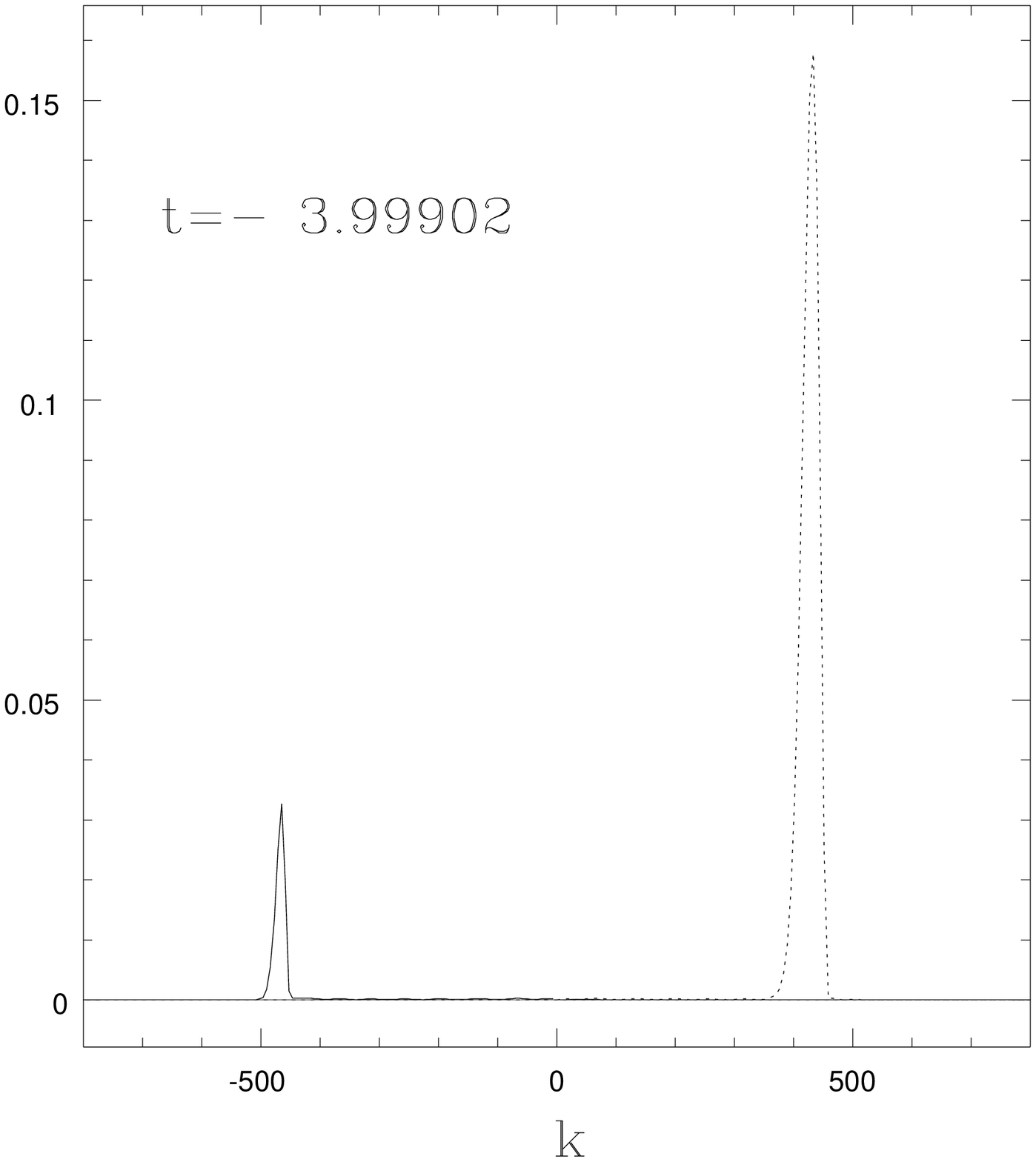,angle=0,height=6.5cm}}
\fcaption{A wavepacket bouncing off the horizon in the Unruh model, taken from
Unruh's numerical integration results. The dispersion relation is
(\ref{FUN}), with $k_0=512$ and $n=1$, and the Hawking temperature
is $T=1.37$.
On the left is the wavepacket at three different times, with time going forward
from bottom to top. The horizon is located at about $x=0.704$, and the black
hole lies to the right.
On the right are power spectra in $k$ of the initial and final wavepackets.}
\label{dumb}
\end{figure}

Unruh's computation\cite{Unruhdumb} starts with a purely outgoing, low
wavevector,
positive Killing frequency wavepacket in the asymptotic region where
$v(x)$ is constant (see top left graph in Fig. \ref{dumb}
for the wavepacket, top right graph for its power spectrum, taken
from Ref. \cite{Unruhdumb}). Numerically integrating the partial differential
equation ({\ref{eom})
backwards in time, he finds that this
wavepacket bounces off the horizon (middle left graph of Fig. \ref{dumb})
and moves back out to the asymptotic
region with only very high wavevector components of both signs (bottom
graphs of Fig. \ref{dumb}).
The sign of the wavevector coincides with the sign of the free-fall
frequency, so this `final' wavepacket (which is actually the {\it initial}
wavepacket going forward in time) possesses a negative norm component
as seen from the power spectrum in the bottom right graph of Fig. \ref{dumb}.
Evaluating this norm Unruh obtains the occupation number (\ref{number})
for the outgoing packet. In all cases, the resulting occupation number agrees,
to the accuracy of the computation, with the thermal prediction of
Hawking. The computation contains alot more information
than just this one number however, since a complete description of the
initial wavepacket is obtained. Since the field equation is {\it linear},
one can simply fourier analyze the initial and final packets and read off
the Bogoliubov coefficients for all wavevectors. Unruh did just this,
and again found agreement with the ideal Hawking effect.

These results have since been confirmed in two further calculations.
In one approach\cite{BMPS}, Brout, Massar, Parentani and Spindel (BMPS)
introduced an analytical approximation scheme in which the field equation near
the horizon is solved by fourier transform. Using this technique, they showed
that the spectrum of emitted radiation is thermal in leading order.
In another approach, Corley and Jacobson\cite{CorlJaco} adopted the
dispersion relation (\ref{F}) and exploited stationarity
of the metric to reduce the PDE to a fourth order ODE for each Killing
frequency, which
was then solved either numerically or, in some cases, analytically.
This technique made it possible to go beyond leading order and compute the
deviations from a perfectly thermal spectrum. These deviations are
relatively small at low frequencies where the Hawking radiation is copius, but
are so large at larger frequencies that they eventually convert the exponential
tail of the Planck spectrum to an increasing function. Nevertheless,
the total flux associated with these higher frequencies is very small.

\subsection{Group velocity reversal and mode conversion}
The behavior of a wavepacket propagated back in time can be understood
qualitatively as follows.
Let $\phi$ = $e^{-i(\omega t - k x)}$, substitute
into the equation of motion, and drop terms containing derivatives
in the free fall velocity.  This is equivalent to using the WKB approximation,
and yields the approximate position dependent dispersion relation
\begin{equation}
(\omega - v(x)k)^{2} = F^{2}(k).
\label{disp'}
\end{equation}
This is just the dispersion relation in the local free-fall frame, since the
free-fall frequency $\o'$ is related to the Killing frequency $\o$ by
\begin{equation}
\o'=\o-v(x)k.
\label{otrans}
\end{equation}
The position-dependent dispersion relation is useful for understanding
the motion of wavepackets that are somewhat peaked in both position
and wavevector. A graphical method we have employed\cite{CorlJaco} is described
below.
The same method was used by BMPS\cite{BMPS}, who also found a Hamiltonian
formulation for the wavepacket propagation using Hamilton-Jacobi theory.

Graphs of both sides of equation (\ref{disp'}) are
 shown in figure \ref{undisp}
\begin{figure}[bh]
\centerline{\psfig{figure=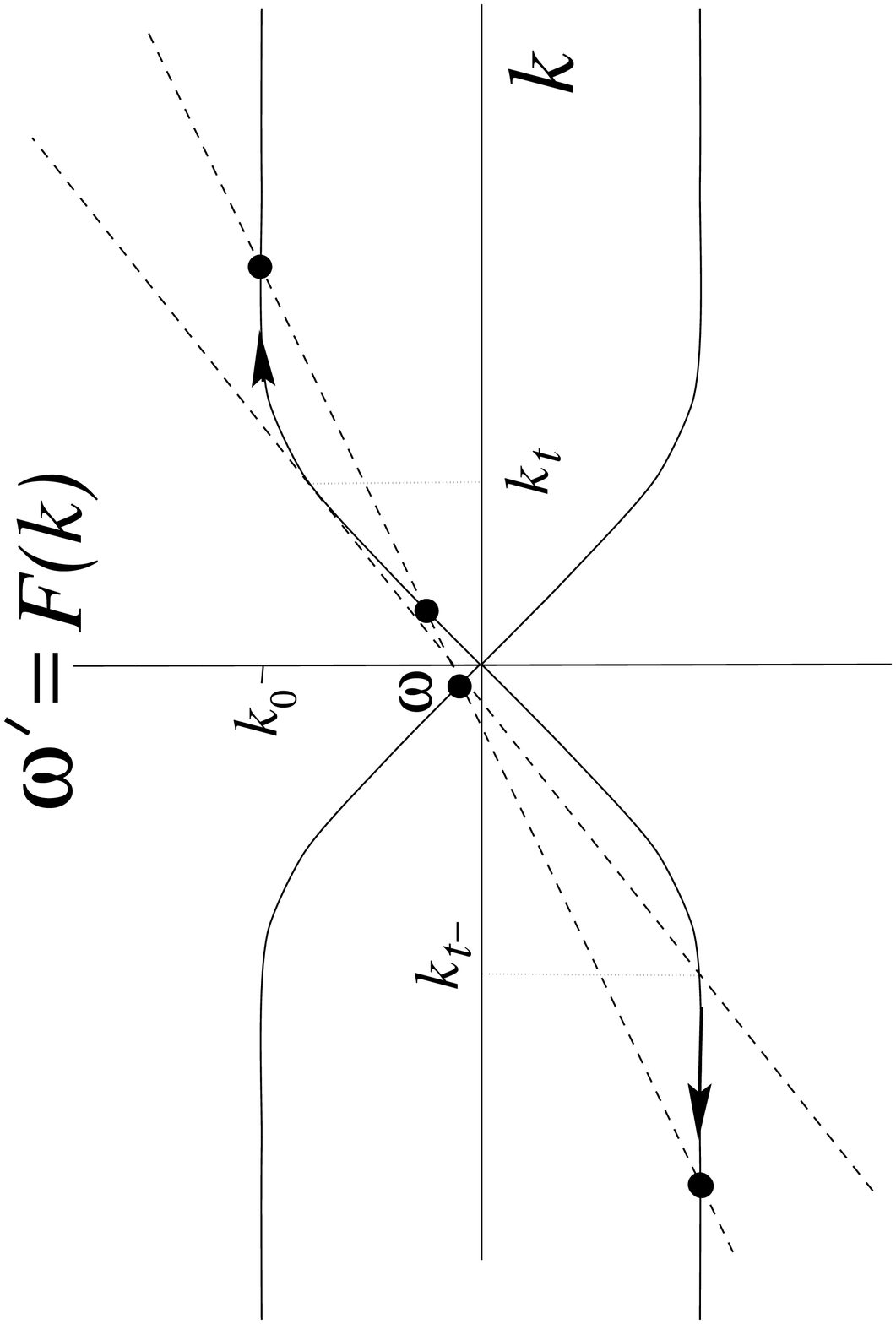,angle=-90,height=7cm}}
\fcaption{Graphical solution of the position-dependent dispersion relation
(\ref{disp'}) with $F(k)$ given by (\ref{FUN}).}
\label{undisp}
\end{figure}
for $F(k)$ given by (\ref{FUN}) and for
two different values of $v$.
As $x$ varies, the slope $-v(x)$ ($=|v(x)|$) of the straight line representing
the left hand side of (\ref{disp'}) changes, but for a given solution
the intercept $\o$ is fixed since the Killing frequency is conserved.
For a given $x$, the
intersection points on the graph correspond to the possible wavevectors
in this approximation. The coordinate velocity
$dx/dt$ of a wavepacket is the group velocity $v_{g} = d\omega/dk$,
which may also be expressed, using (\ref{otrans}),
as $v_{g} = d\omega '/dk + v(x)$.
Note that $v_{g}' \equiv d\omega '/dk$ is the group velocity
of the packet in the free-fall frame, and corresponds to the slope
of the curved line in figure \ref{undisp}.

Now assume the free fall velocity function asymptotes (at large positive $x$)
to a value $v_{o}$ satisfying $-1<v_{o}<0$, and  consider a narrow
wavepacket located far from the hole, centered about
frequency $\omega$, and containing only $k$ values around the smaller
positive root (intersection point) shown in the figure. This is an outgoing
wavepacket, since $d\omega '/dk>|v|$. Therefore, going
backwards in time, the packet moves towards the hole.  As $x$ decreases
$|v(x)|$ increases, so the slope of the straight line increases, until
eventually the straight line becomes tangent to the dispersion curve.
At this point $v_{g}$ drops to zero.  If $\omega$ is very small compared to
$k_0$, then this stopping point $x_t$ occurs when $v(x)$ is very close to $-1$,
that is, just barely outside the horizon.

What happens at the stopping point? It was incorrectly suggested in
Ref. \cite{Jacobson1} that
the wavepackets just asymptotically approach limiting position $x_t$ and
wavevector $k_t$. However, at the stopping point, the simple picture
of a wavepacket peaked around a single wavevector has broken down.
The wave packet has a significant spread in wavevectors which fail to satisfy
the local dispersion relation. Furthermore, in this region there are nearby
solutions to the dispersion relation. Just outside $x_t$, the straight
line cuts the dispersion curve in two nearby points which approach each other
as $x$ approaches $x_t$. Simple WKB analysis is unable to say what happens.
But a supplementary consideration gives the answer.
As pointed out by Unruh\cite{Unruhpc}, this is an unstable situation: for $k$
slightly above $k_t$ the group velocity drops below zero (i.e. the
comoving group velocity drops below the magnitude of the free-fall velocity)
so, backwards in time,  the wavepacket tends to move back {\it away}
from the horizon. Once this begins to happen, $k$ continues to increase
as the wavepacket moves further away. Exactly this behavior was found in
Unruh's numerical solution\cite{Unruhdumb} to the PDE.

This is not the end of the story however. There is yet another
``nearby" solution to the dispersion relation, on the negative wavevector
branch, which is excited as well.
This process is an example of ``mode conversion"\cite{Stix,Swanson}.
The mode conversion is strong if the wavelength and phase velocity of one mode
is close to that of another. It appears that
this condition for strong conversion is not met, since the negative $k$
solution $k_{t-}$ to (\ref{disp'}) at $x_t$ is not very close in magnitude to
$k_t$ (see Fig. \ref{undisp}.) However, one must remember that the wavepacket
here has a significant spread in wavevectors which fail to satisfy the local
dispersion relation.
Thus the wavepacket can pick up other modes that fulfill the conversion
criterion. That the wavepacket {\it must}
pick up a negative wavevector piece is clear in the case of the ordinary
wave equation, since there causality implies that the wavepacket strictly
vanishes inside the horizon, which cannot be accomplished with only
positive wavevectors. In the Unruh model, the wavepacket satisfies
approximately the ordinary wave equation until it gets very close to the
horizon, so one should expect that a similar amount of mode conversion
takes place. Evidently the negative wavevector mode
{\it does} couple in strongly for sufficiently small $\o$, as shown both by
Unruh's solution of the PDE and
by the ODE methods applied by BMPS\cite{BMPS} and ourselves\cite{CorlJaco}.
The ``converted", negative
wavevector, wavepacket also has a negative group velocity and so also
moves, backwards in time, away from the hole.
The end result thus consists of two wavepackets, one constructed of large
positive $k$ wavevectors and the other of large negative
$k$ wavevectors, both propagating away from the hole
and reaching the asymptotically flat (constant free fall velocity) region.
Note that the negative wavevector component will consist of wavevectors
of magnitude  slightly larger than the positive wavevector component.
This is clear from the position of the intersection points in Fig.
\ref{undisp},
and is borne out in Unruh's computation (Fig. \ref{dumb}) as well.
The number of created particles in the final, late time, wavepacket is
given by the norm of the negative wavevector part of the initial,
early time wavepacket.

\section{The stationarity puzzle}
\label{puzzle}

Particle production via the Hawking effect in the case of a
black hole that forms in a collapse process is ordinarily understood
to depend critically on the fact that the metric is not static for all times.
Although the radiating modes come in from infinity and go back out to infinity,
their Killing frequency is not conserved because they propagate through the
time dependent part of the spacetime.
The way the usual Hawking effect transpires in a strictly stationary spacetime
is that the outgoing wavepackets are traced backwards to parts that do
{\it not}
make it back out to infinity, but rather cross the white hole horizon, at which
point the Unruh boundary condition on the quantum state is imposed. The piece
of the wavepackets that scatters off the curvature and does make it back out to
infinity is not associated with particle creation. If, as in the Unruh
model, the entire wavepacket turns around and goes back out, remaining for
all time in the stationary region of the spacetime,
then it would seem that there can be no particle creation at all. So how does
the Unruh model yield a nontrivial Hawking flux?

The first answer is that the wavepackets have not been followed
all the way out to infinity. Another way to put it is that we
have taken the free fall frame, in which the boundary condition is imposed, to
be moving towards the black hole at infinity, rather than coinciding with the
rest frame of the black hole at infinity. (Technically, this corresponds to
the fact that the metric function $v(x)$ does not vanish at infinity.)
But what happens if we take $v(x)$ to vanish at infinity, and continue to
follow the wavepackets backwards in time, not stopping to impose the quantum
state boundary condition until the packets reach infinity---or do something
else?
Do the wavepackets propagating backwards in time ever
reach infinity? Is there any Hawking radiation?

What happens in the Unruh model with the dispersion relation
(\ref{FUN}) is that the magnitude of the wavevector grows without bound as the
wavepacket moves outward where $v(x)$ is falling to zero.
Thus, even though the difference between the free-fall and Killing frames is
going to zero, the wavevector is diverging in such a way that the
wavepacket always maintains a negative free-fall frequency part of the same,
negative, norm. Thus the Hawking effect indeed occurs.
{}From this analysis we
see that the Unruh model, while it entails a strict cutoff in free-fall
frequency, involves in an essential way arbitrarily high wavevectors, i.e.,
arbitrarily short wavelengths.  Insofar as we wish to explore the consequences
of a fundamental short distance cutoff on the Hawking effect, this is an
unsatisfactory feature of the model.
The outgoing modes emerging from the black hole region still arise from
arbitrarily short wavelength modes. Using the dispersion relation
(\ref{F}) on the other hand, the wavevectors are bounded by $k_0$.
In this case something very strange seems to happen\cite{CorlJaco},
which remains to be
understood: the positive norm piece of the wavepacket goes backwards
in time out to infinity at superluminal group velocity, and the velocity of the
negative norm piece goes past positive infinity to negative infinity, which
seems to result in a propagation ``back to the future".

In search of a more realistic model, it is interesting to go back and consider
the
behavior of a wavepacket propagated backwards in time using the dispersion
relation of liquid helium-4 (see Fig. \ref{disps}). Up to the reversal of group
velocity outside
the horizon, the behavior is the same as for the Unruh model.
After that (i.e. {\it before} that) the packet will go over
the first maximum of the dispersion curve, at which point its comoving group
velocity changes sign, which only pushes it away from the horizon even faster.
Eventually, however, it approaches another turn around point, near the
roton minimum, where the
free-fall frequency line becomes tangent once again to the dispersion
curve. It seems reasonable to suppose that what happens here is another
reversal
of direction and further mode conversion: the wavepacket continues along the
dispersion curve, falling
back towards the horizon, until it runs off the end of the quasi-particle
spectrum, where it is presumably unstable and has a wavelength on the order
of the interatomic spacing. Or, perhaps, the instability sets in even before
this point. It thus appears that there is no way to analyze
the helium model fully at the level of quasi-particle field theory. Rather, the
many particle dynamics must be directly confronted. It would be very
interesting, though perhaps very difficult, to analyze this many body problem.

\section{Acknowledgements}

These notes are based on lectures given at the First Mexican School
on Gravitation and Mathematical Physics, Guanajuato, December 1994.
I am grateful to the organizers for the opportunity to
present these lectures. I am also grateful to my collaborator
S. Corley for countless discussions, and to W.G. Unruh for answering
my many questions. This work was supported in part by NSF grant PHY94-13253.

\end{document}